\documentclass[twocolumn]{aastex63}
\usepackage[utf8]{inputenc}
\usepackage{graphicx}
\usepackage{amssymb}
\usepackage{amsmath}
\usepackage{float}
\usepackage{multirow}
\usepackage{textcomp}
\usepackage{gensymb}
\usepackage{hyperref}
\usepackage{natbib}
\usepackage{comment}
\usepackage{systeme}
\usepackage[Symbol]{upgreek}
\usepackage{xcolor}

\shorttitle{High-resolution low-frequency observations of Abell 2256}
\shortauthors{K. Rajpurohit et al.}

\begin{document}

\title{Deep low-frequency radio observations of Abell 2256 I: The filamentary radio relic}
\author[0000-0001-7509-2972]{K. Rajpurohit}
\affil{DIFA - Universit\'a di Bologna, via Gobetti 93/2, 40129 Bologna, Italy}
\affil{INAF-IRA, Via Gobetti 101, 40129 Bologna, Italy} 
\affil{Th\"uringer Landessternwarte, Sternwarte 5, 07778 Tautenburg, Germany}
\author[0000-0002-0587-1660]{R. J. van Weeren}
\affil{Leiden Observatory, Leiden University, PO Box 9513, 2300 RA Leiden, The Netherlands}

\author{M. Hoeft}
\affil{Th\"uringer Landessternwarte, Sternwarte 5, 07778 Tautenburg, Germany}

\author[0000-0002-2821-7928]{F. Vazza}
\affil{DIFA - Universit\'a di Bologna, via Gobetti 93/2, 40129 Bologna, Italy}
\affil{INAF-IRA, Via Gobetti 101, 40129 Bologna, Italy} 
\affil{Universit\"at Hamburg, Hamburger Sternwarte, Gojenbergsweg 112, 21029, Hamburg, Germany}

\author[0000-0003-4120-9970]{M. Brienza}
\affil{DIFA - Universit\'a di Bologna, via Gobetti 93/2, 40129 Bologna, Italy}
\affil{INAF-IRA, Via Gobetti 101, 40129 Bologna, Italy} 

\author[0000-0002-9478-1682]{W. Forman}
\affil{Harvard-Smithsonian Center for Astrophysics, 60 Garden Street, Cambridge, MA 02138, USA}

\author{D. Wittor}
\affil{Universit\"at Hamburg, Hamburger Sternwarte, Gojenbergsweg 112, 21029, Hamburg, Germany}

\author[0000-0001-7058-8418]{P. Dom\'{i}nguez-Fern\'{a}ndez}
\affil{Department of Physics, School of Natural Sciences UNIST, Ulsan 44919, Korea}

\author{S. Rajpurohit}
\affil{Molecular Foundry, Lawrence Berkeley National Laboratory, Berkeley, CA 94720, USA}

\author{C. J. Riseley}
\affil{DIFA - Universit\'a di Bologna, via Gobetti 93/2, 40129 Bologna, Italy}
\affil{INAF-IRA, Via Gobetti 101, 40129 Bologna, Italy} 
\affil{CSIRO Space \& Astronomy, PO Box 1130, Bentley, WA 6102, Australia;}

\author[0000-0002-9325-1567]{A. Botteon}
\affil{Leiden Observatory, Leiden University, PO Box 9513, 2300 RA Leiden, The Netherlands}

\author [0000-0002-5815-8965]{E. Osinga}
\affil{Leiden Observatory, Leiden University, PO Box 9513, 2300 RA Leiden, The Netherlands}

\author{G. Brunetti}
\affil{INAF-IRA, Via Gobetti 101, 40129 Bologna, Italy}

\author{E. Bonnassieux}
\affil{DIFA - Universit\'a di Bologna, via Gobetti 93/2, 40129 Bologna, Italy}
\affil{INAF-IRA, Via Gobetti 101, 40129 Bologna, Italy}

\author{A. Bonafede}
\affil{DIFA - Universit\'a di Bologna, via Gobetti 93/2, 40129 Bologna, Italy}
\affil{INAF-IRA, Via Gobetti 101, 40129 Bologna, Italy} 
\affil{Universit\"at Hamburg, Hamburger Sternwarte, Gojenbergsweg 112, 21029, Hamburg, Germany}

\author{A. S. Rajpurohit}
\affil{Astronomy \& Astrophysics Division, Physical Research Laboratory, Ahmedabad 380009, India}

\author{C. Stuardi}
\affil{DIFA - Universit\'a di Bologna, via Gobetti 93/2, 40129 Bologna, Italy}
\affil{INAF-IRA, Via Gobetti 101, 40129 Bologna, Italy} 

\author[0000-0003-2792-1793]{A. Drabent}
\affil{Th\"uringer Landessternwarte, Sternwarte 5, 07778 Tautenburg, Germany}

\author[0000-0002-3369-7735]{M. Br{\"u}ggen}
\affil{Universit\"at Hamburg, Hamburger Sternwarte, Gojenbergsweg 112, 21029, Hamburg, Germany}

\author{ D. Dallacasa}
\affil{DIFA - Universit\'a di Bologna, via Gobetti 93/2, 40129 Bologna, Italy}
\affil{INAF-IRA, Via Gobetti 101, 40129 Bologna, Italy} 

\author[0000-0001-5648-9069]{T.W. Shimwell}
\affil{ASTRON, Netherlands Institute for Radio Astronomy, Oude Hoogeveensedijk 4, 7991 PD, Dwingeloo, The Netherlands}
\affil{Leiden Observatory, Leiden University, PO Box 9513, 2300 RA Leiden, The Netherlands}

\author[0000-0001-8887-2257]{H.J.A. R{\"o}ttgering}
\affil{Leiden Observatory, Leiden University, PO Box 9513, 2300 RA Leiden, The Netherlands}

\author{F. de Gasperin}
\affil{Universit\"at Hamburg, Hamburger Sternwarte, Gojenbergsweg 112, 21029, Hamburg, Germany}

\author{G. K. Miley}
\affil{Leiden Observatory, Leiden University, PO Box 9513, 2300 RA Leiden, The Netherlands}

\author{M. Rossetti}
\affil{INAF - IASF Milano, via A. Corti 12, 20133 Milano, Italy}

\correspondingauthor{Kamlesh Laxmi Rajpurohit}
\email{kamlesh.rajpurohit@unibo.it}


\begin{abstract}
We present deep and high-fidelity images of the merging galaxy cluster Abell 2256 at low frequencies using the upgraded Giant Metrewave Radio Telescope (uGMRT) and LOw-Frequency ARray (LOFAR). This cluster hosts one of the most prominent known relics, with a remarkably spectacular network of filamentary substructures. The new uGMRT (300-850\,MHz) and  LOFAR (120-169\,MHz) observations, combined with the archival Karl G. Jansky Very Large Array (VLA; 1-4\,GHz) data, allowed us to carry out the first spatially resolved spectral analysis of the exceptional relic emission down to $6\arcsec$ resolution over a broad range of frequencies. Our new sensitive radio images confirm the presence of complex filaments of magnetized relativistic plasma  also at low frequencies. We find that the integrated spectrum of the relic is consistent with a single power law, without any sign of spectral steepening at least below 3\,GHz. Unlike previous claims, the relic shows an integrated spectral index of $-1.07\pm0.02$ between 144\,MHz and 3\,GHz, which is consistent with the (quasi)stationary shock approximation. The spatially resolved spectral analysis suggests that the relic surface very likely traces the complex shock front, with a broad distribution of Mach numbers propagating through a turbulent and dynamically active intracluster medium. Our results show that the northern part of the relic is seen edge-on and the southern part close to face-on. We suggest that the complex filaments are regions where higher Mach numbers dominate the (re-)acceleration of electrons that are responsible for the observed radio emission.
\end{abstract}


\section{Introduction}
 \label{sec:intro}

Galaxy clusters undergoing mergers often show spectacular, megaparsec-scale radio relics and radio halos \citep[see][for a recent review]{vanWeeren2019}. The radio spectra of such sources are steep ($\alpha \leq-1.0$ where $S_{\nu}\propto\nu^{\alpha}$, with spectral index $\alpha$). Cluster-scale radio sources are powered by the dissipation of kinetic energy during cluster formation process, although the particle acceleration detail is not yet fully understood \citep[e.g.,][]{Brunetti2014}. 

Radio relics are elongated radio sources typically found in the outskirts of merging galaxy clusters. They are usually strongly polarized at gigahertz frequencies \citep[e.g.,][]{Bonafede2012,vanWeeren2010,vanWeeren2012a,Kale2012,Owen2014,Stuardi2019,DiGennaro2021,Rajpurohit2020b,Rajpurohit2021d} and believed to originate from shock fronts generated in the intracluster medium (ICM) when galaxy clusters merge \citep{Ensslin1998,Roettiger1999}. This connection has been established by finding X-ray surface brightness and temperature discontinuities at the location of some relics \citep[see e.g.,][]{Sarazin2013,Ogrean2013,Shimwell2015,vanWeeren2016a,Botteon2016,Tholken2018,Gennaro2019}.

\setlength{\tabcolsep}{10pt}
\begin{table*}[!htbp]
\caption{Observational overview: uGMRT, LOFAR, and VLA observations.}
\begin{center}
\begin{tabular}{ l  c  c c c  c c}
  \hline  \hline  
& uGMRT Band\,3  &  uGMRT Band\,4  &LOFAR HBA  $^{_\dagger}$ & VLA L-band $^{\ddagger}$  & VLA S-band $^{\ddagger}$  \\  
  \hline  
Frequency range&300-500\,MHz&550-950\,MHz &120-169 MHz&1-2\,GHz, &2-4\,GHz\\ 
Channel width & 97\,kHz & 49\,kHz &12.2\,kHz&1\,MHz &2\,MHz\\ 
No. of spectral window &1 &1 &-&16 &16\\ 
No. of channels &4096 &4096 &-&64 &64\\ 
On source time &10\,hr &8\,hr &16\,hr&24\,hr &6\,hr  \\
LAS$^{\ast}$ &$1920\arcsec$&1020\arcsec &$3.8\degree$&970\arcsec&490\arcsec\\ 
\hline 
\end{tabular}
\end{center}
{Notes. Full Stokes polarization information was recorded for the uGMRT Band\,4 and VLA L-band, S-band; $^{\dagger}$ for data reduction of the LOFAR 144\,MHz observations we refer to Osinga et al. in prep.; $^{\ddagger}$archival VLA L and S-band data; $^{\ast}$Largest angular scale (LAS) that can be recovered with the mentioned observations. }
\label{Tabel:obs}
\end{table*} 

Current relic formation scenarios, namely, shock acceleration, shock re-acceleration, and shock compression, differ in the predictions of the morphological and spectral characteristics of relics. In the shock acceleration scenario, it is believed that the kinetic energy dissipated by shocks powers the radio emission via diffusive shock acceleration (DSA) of thermal electrons in the ICM \citep[e.eg.,][]{Ensslin1998,Blasi1999,Dolag2000,Hoeft2007}. This model predicts a power-law energy spectrum for the cosmic ray electrons (CRe). However, there are a few relics that show evidence of steepening at high frequencies above $1\,\rm GHz$ \citep{Trasatti2015,Malu2016}, and a flatter spectral index (above $-1$) that is incompatible with the shock acceleration scenario \citep[e.g.,][]{Trasatti2015}. Most importantly, a power-law CRe energy distribution ranging from thermal energies to those relevant for the synchrotron emission may require an unphysical acceleration efficiency at the shock front and thus cannot explain the high radio luminosities observed in some relics \citep[e.g.,][]{Brunetti2014,Vazza2014,Vazza2015,Botteon2020a}. Therefore, the main remaining debate regarding the origin of relics is whether pre-acceleration (i.e., a population of mildly-relativistic fossil electrons) is needed to explain the origin of at least some relics.

In the re-acceleration scenario, the shock front reaccelerates electrons via DSA from an existing population of mildly relativistic electrons \citep{Markevitch2005,Kang2016a}. There are a few examples that appear to show a connection between lobes of active galactic nuclei (AGN), i.e., they provide evidence that the shock fronts re-accelerate CRe of a fossil population \citep[e.g.,][]{Bonafede2014,vanWeeren2017a,Stuardi2019,Shimwell2015,HyeongHan2020}. This mechanism alleviates the efficiency problem \citep[e.g.][]{2013MNRAS.435.1061P,ka12}. If the fossil population is homogeneously distributed throughout the relic or the shock Mach number is strong, the re-acceleration scenario also predicts a power-law spectrum. However, if the fossil plasma in the ICM is not continuously distributed, a high frequency spectral steepening in the overall spectrum is expected. 

In the shock compression model, relics are produced by adiabatic compression of fossil radio plasma by the passage of a shock front in the ICM \citep{Ensslin2001}. This model predicts a spectral break at high frequencies since the compression cannot boost electrons at very high energies.

Radio relics often display an irregular surface brightness, and recent high-resolution observations provided evidence of fine filaments on various scales, for instance, in Abell 2256 \citep{Owen2014}, 1RXS\,J0603.3+4214 \citep[aka the Toothbrush relic,][]{Rajpurohit2018,Rajpurohit2020a}, CIZA\,J2242.8+5301 \citep[aka the Sausage relic,][]{Gennaro2018}, Abell 2255 \citep{Botteon2020}, and  MACS\,J0717.5+3745 \citep{vanWeeren2017b}. The origin of these filamentary structures is unknown.

In this paper, we present the results of deep low frequency radio observations of the galaxy cluster Abell 2256 with the upgraded Giant Metrewave Radio Telescope (uGMRT) and the new LOw-Frequency ARray \citep[LOFAR;][]{Haarlem2013} High Band Antenna (HBA) observations (Osinga et al. in prep). These observations were mainly undertaken to provide higher-resolution radio images of the  relic in Abell 2256, thus allowing us to study the exceptional radio emission in more detail than had been done previously over a large frequency coverage. We also use the archival Karl G. Jansky Very Large Array (VLA) in L and S-band.

The main aim of this paper is to study the origin of the large relic in Abell 2256. We attempt to test shock acceleration models for this relic. We also aim to find whether there is a high frequency spectral break in the integrated spectrum between 144\,MHz and 3\,GHz. The question of whether or not such a spectral break exists is essential, since it sheds light on the mechanisms which generate radio relics. Detailed spectral analysis allows us to distinguish between different  models (shock acceleration and re-acceleration). 

The layout of this paper is as follows. In Sect.\,\ref{observations}, we present an overview of the observations and data reduction. The new radio images are presented in Sect.\,\ref{results}. The results obtained with the spectral analysis are described in Sects.\,\ref{relic_analysis}-\ref{offset_dis}, followed by a summary in Sect.\,\ref{summary}.

\setlength{\tabcolsep}{13.0pt}   
 \begin{table*}[!htbp]
\caption{Image properties}
\begin{center}
\begin{tabular}{c c c r c c c r}
\hline\hline
   & Name & Restoring Beam & Robust  & \textit{uv}-cut & \textit{uv}-taper & RMS noise\\ 
&&&parameter&&&$\upmu\,\rm Jy\,beam^{-1}$\\
\hline
\hline
&IM1&$6\arcsec \times 6\arcsec$&$-0.5$&$ \geq\rm0.1\,k\uplambda$&$-$&97\\
LOFAR  &IM2&$10\arcsec \times 10\arcsec$&$-0.5$&$\geq\rm0.1\,k\uplambda$&10\arcsec&110\\
 &IM3&$20\arcsec \times 20\arcsec$&$-0.5$&$ \geq\rm0.1\,k\uplambda$&16\arcsec&191 \\
  \hline 
&IM4&$3\arcsec \times 3\arcsec$&$-0.6$&$-$&$-$&7\\
&IM5&$5\arcsec \times 5\arcsec$&0.0&$-$&$-$&4\\
uGMRT Band\,4&IM6&$6\arcsec \times 6\arcsec$&$-0.5$&$ \geq\rm0.1\,k\uplambda$&3\arcsec&9\\
&IM7&$10\arcsec \times 10\arcsec$&$-0.5$&$ \geq\rm0.1\,k\uplambda$&10\arcsec&12\\
&IM8&$20\arcsec \times 20\arcsec$&0.0&$-$&15\arcsec&18\\
 \hline 
&IM9&$8\arcsec \times 8\arcsec$&$-0.2$&$-$&$-$&36\\
uGMRT Band\,3 &IM10&$10\arcsec \times 10\arcsec$&$-0.5$&$ \geq\rm0.1\,k\uplambda$&10\arcsec&34\\
&IM11&$20\arcsec \times 20\arcsec$&0.0&$-$&15\arcsec&81\\
\hline   
 &IM12&$5\arcsec \times 5\arcsec$&0.0&$-$&4\arcsec&6\\
&IM13&$6\arcsec \times 6\arcsec$&$-0.5$&$ \geq\rm0.1\,k\uplambda$&3\arcsec&8\\
VLA L-band&IM14&$10\arcsec \times 10\arcsec$&$-0.5$&$ \geq\rm0.1\,k\uplambda$&10\arcsec&8\\
&IM15&$20\arcsec \times 20\arcsec$&0.0&$-$&15\arcsec&11\\
\hline   
VLA S-band&IM16&$6\arcsec \times 6\arcsec$&0.0&$-$&$-$&5\\
 &IM17&$10\arcsec \times 10\arcsec$&$-0.5$&$ \geq\rm0.1\,k\uplambda$&10\arcsec&9 \\
 \hline  
 \end{tabular}
 \end{center}
{Notes. Imaging was always performed in {\tt WSCLEAN} using {\tt multiscale} and with {\tt Briggs} weighting scheme}. 
\label{imaging}
\end{table*}


\section{Abell 2256}
\label{target}
Nearby cluster ($z=0.058$) Abell 2256 contains a plethora of complex radio sources, see Fig.\,\ref{overlay}.  The cluster has one of the richest variety of radio structures of any known cluster \citep{Bridle1976,Bridle1979,Rottgering1994,Clarke2006,vanWeeren2009a,Brentjens2008,Kale2010,vanWeeren2012b,Owen2014,Trasatti2015}. One of the most intriguing radio source in the field is the large, complex filamentary relic (R), which shows many long pronounced filaments stretching across its entire structure \citep{Owen2014} and making it unlike other known relics.

The relic shows a high degree of polarization and large-scale magnetic field ordering, as well as significant rotation measure (RM) fluctuations \citep{Owen2014,Ozawa2015}. This relic is the one notable exception as it shows a spectral index of about $-0.85\pm0.01$ \citep{vanWeeren2012b,Trasatti2015} between 63\,MHz and 1.4\,GHz. Moreover, a break in the spectrum below 1.4\,GHz has also been reported  \citep{Trasatti2015}.

In DSA test-particle approximation, the shock strength is constant in the cooling time of the CRe relevant for the observation. The integrated spectral index ($\alpha_{\rm int}$) is then 0.5 steeper than the injection index ($\alpha_{\rm inj}$), 
\begin{equation}
\alpha_{\rm int}=\alpha_{\rm inj}-0.5.
\end{equation}
The majority of relics follow this approximation. The flat spectral index of the relic in Abell 2256 is one of the few examples where the observations are in clear discord with the shock acceleration model. However, previous studies \citep{vanWeeren2012b,Trasatti2015} of the integrated spectrum of the relic in Abell 2256 were not performed with matching \textit{uv}-coverage (i.e., different ranges of  spatial scales were recovered at different frequencies). Therefore, it remains unclear whether the spectrum of the relic steepens above 1.4\,GHz or not. Recently, it has been found that the integrated spectrum of the Toothbrush \citep{Rajpurohit2020b} and the Sausage relics \citep{Loi2020} both follow a close power-law between 150\,MHz to 18.6\,GHz and has spectral indices steeper than $-1$ contrary to previous claims \citep{Stroe2016,Kierdorf2016}.

The cluster is also known to host a steep spectrum radio halo and several complex radio sources (A, B, C, F1, F2, F3, and I), see Fig.\,\ref{overlay} for labeling \citep{Rottgering1994,Intema2009,Clarke2006,Brentjens2008,vanWeeren2009,vanWeeren2012b,Owen2014}. 

At X-ray wavelengths, the cluster is luminous \citep[$L_{\rm X,0.1-2.4\,\rm keV}=3.7\times 10^{44}\,\rm erg\,s^{-1}$;][]{Ebeling1998}. Recent X-ray studies provide evidence for five X-ray surface brightness discontinuities: three cold fronts and two shock fronts \citep{Ge2020,Breuer2020}. One of the shock fronts, which has a Mach number $\mathcal{M}=1.26\pm0.06$, is detected at the northwest of the large relic \citep{Ge2020}. However, there is an offset of about 150\,kpc between the relic emission at 1.5\,GHz and the detected X-ray density/temperature jump. 

The cluster has a total mass of about $\rm M_{500}=(6.11\pm0.40)\times10^{14}\,M_{\odot}$ \citep{Planck2011}. Optical studies of the galaxy distribution reveal that the cluster consists of three separate components \citep{Berrington2002,Miller2003}. Both optical and X-ray observations provide strong evidence that Abell 2256 is undergoing a merger event between a main cluster component, a major sub-component and a third infalling group \citep{Briel1991,Briel1994,Sun2002,Berrington2002,Miller2003,Ge2020}.

Throughout this paper, we adopt a flat $\Lambda$CDM cosmology with $H_{\rm{ 0}}=69.6$ km s$^{-1}$\,Mpc$^{-1}$, $\Omega_{\rm{ m}}=0.286$, and $\Omega_{\Lambda}=0.714$. At the cluster's redshift, $1\arcsec$ corresponds to a physical scale of 1.13\,kpc. All output images are in the J2000 coordinate system and are corrected for primary beam attenuation. 


\section{Observations and data reduction}
\label{observations}
\subsection{uGMRT}
Abell 2256 was observed with the upgraded GMRT in Band\,4 (project code 36\_006) and Band\,3 (project code 39\_019) using the GMRT Wideband Backend (GWB). The observations were carried out on 2019 September 9 and  2020 November 20 in Band\,4 and Band\,3 respectively. In Table\,\ref{Tabel:obs}, we summarize the observational details.  Sources 3C\,48 and 3C\,286 were included as flux calibrators.

The wideband GMRT data were processed using the Source Peeling and Atmospheric Modeling \citep[$\tt{SPAM}$;][]{Intema2009}, pipeline{\footnote{\url{http://www.intema.nl/doku.php?id=huibintemaspampipeline}}}. The $\tt{SPAM}$ pipeline performs automated rounds of both direction-independent and direction-dependent calibration and imaging, as well as radio frequency interference (RFI) flagging. For details about the main data reduction steps, we refer to \cite{Rajpurohit2021c}. In summary, the flux densities of the primary calibrators were set according to \citet{Scaife2012}. Following flux density scale calibration, the data were averaged, flagged and corrected for the bandpass. To correct the phase gains of the target field, we started from a global sky model obtained with the GMRT narrow band data. To produce deep, full continuum images, the calibrated sub-bands were combined. The final deconvolution was performed in {\tt WSClean} \citep{Offringa2014} using $\tt{multiscale}$ and $\tt{Briggs}$ weighting with robust parameter 0.


\begin{figure*}[!thbp]
    \centering
    \includegraphics[width=1.00\textwidth]{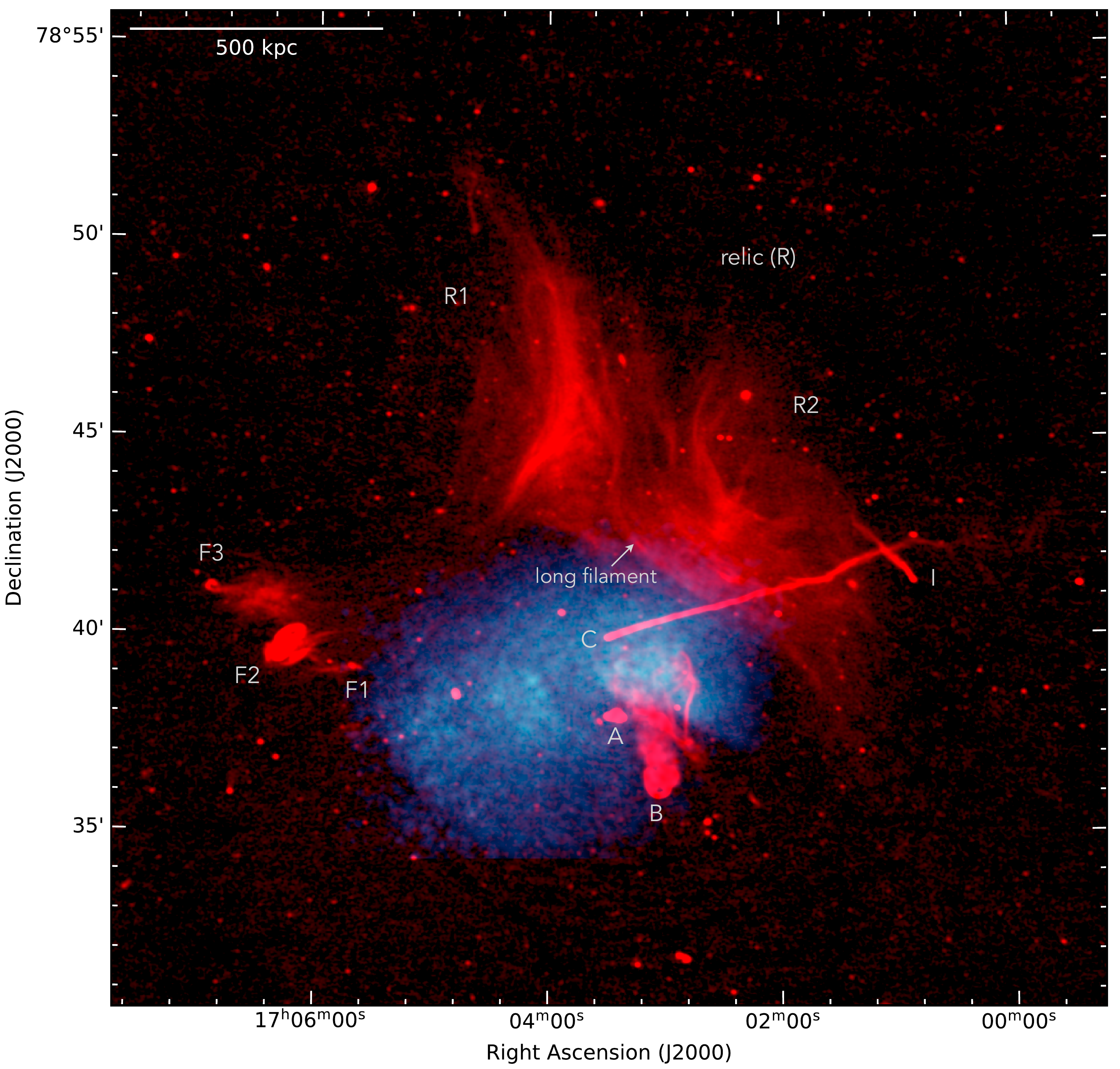}
    \vspace{-0.5cm}
 \caption{uGMRT (550-850 MHz) and \textit{Chandra} $0.5-2.0$\,keV band X-ray overlay of the galaxy cluster Abell 2256. The intensity in red shows the radio emission observed with uGMRT Band\,4 at a central frequency of 675\,MHz. The intensity in blue shows \textit{Chandra} X-ray emission smoothed to $3\arcsec$. The image shows a large relic (R) with complex filaments on various scales, as well as numerous other features. The image properties are given in Table\,\ref{imaging}, IM5.}
      \label{overlay}
  \end{figure*}

\subsection{LOFAR}
The cluster was observed with the LOFAR HBA in 2018 and 2019. The Abell 2256 field was covered in two different observing sessions (project codes LC9\_008 and LT10\_10). These observations were conducted in HBA dual inner mode. A detailed description of the observations, data reduction procedure, and images will be provided in Osinga et al. in prep. To summarize, data reduction and calibration was performed with the LoTSS DR2 pipeline \citep{Tasse2020} followed by a final ``extraction+self-calibration" scheme \citep{vanWeeren2020}.

\subsection{VLA}
We reduced VLA archival L- and S-band data obtained with wideband receivers. These observations are first presented in \citet{Owen2014} and \citet{Ozawa2015}. The L-band observations were taken in ABCD configurations (project codes 10B-154 and 12B-120) and the S-band in the C-configuration (project code 13A-131). For observational details see Table\,\ref{Tabel:obs}. All four polarization products (RR, RL, LR, and LL) were recorded. In the L-band, for each configuration 3C\,48 was included as the primary calibrator and observed for 5-10 minutes at the start of the observing run or, in some cases, at the end. 3C\,286 was a primary calibrator in S-band. The radio source J1800$+$7828 was included as a phase calibrator for both S- and L-band observations.

The data were calibrated and imaged with the Common Astronomy Software Applications \citep[$\tt{CASA}$;][]{McMullin2007} package. Data obtained from different observing runs were calibrated separately but in the same manner. The first step of data reduction consisted of Hanning smoothing of the data. The data were then inspected for RFI, and the affected data were flagged using ${\tt tfcrop}$ mode from the ${\tt flagdata}$ task. Low-amplitude RFI was flagged using ${\tt AOFlagger}$ \citep{Offringa2010}. Following flagging, we determined and applied elevation-dependent gain tables and antenna offset positions.  Next, we corrected for the bandpass using the calibrator 3C\,48. This prevents flagging of good data due to the bandpass roll-off at the edges of the spectral windows.


We used the S-band and L-band 3C48, 3C286, and 3C138 models provided by the ${\tt CASA}$ software package and set the flux density scale according to \cite{Perley2013}. An initial phase calibration was performed using both the calibrators over a few channels per spectral window. We corrected for the parallel-hands (RR and LL) antenna delays and determined the bandpass response using the calibrator  3C48. Applying the bandpass and delay solutions, we proceeded with the gain calibration for the primary calibrators.  All relevant calibration solutions were then applied to the target field. For all different observing runs, the resulting calibrated data were averaged by a factor of 4 in frequency per spectral window. 

\subsection{Flux density scale}

The overall flux scale for all observations (LOFAR, uGMRT, and VLA) was checked by comparing the spectra of compact sources in the field of view between 144\,MHz and 4\,GHz.  The LOFAR HBA flux densities were found to be slightly high. Therefore, we use a scaling factor of 0.83 to correct the flux density scale of the LOFAR HBA maps.

The uncertainty in the flux density measurements was estimated as
\begin{equation}
\Delta S =  \sqrt {(f \cdot S)^{2}+{N}_{{\rm{ beams}}}\ (\sigma_{{\rm{rms}}})^{2}},
\end{equation}
where $f$ is the absolute flux density calibration uncertainty, $S$ is the flux density, $\sigma_{{\rm{ rms}}}$ is the RMS noise, and $N_{{\rm{beams}}}$ is the number of beams. We assume absolute flux density uncertainties of 10\,\% for uGMRT Band\,3 \citep{Chandra2004} and LOFAR data, 5\% for uGMRT Band\,4, and 2.5\,\% for the VLA data. 


\begin{figure*}[!thbp]
    \centering
    \includegraphics[width=1.00\textwidth]{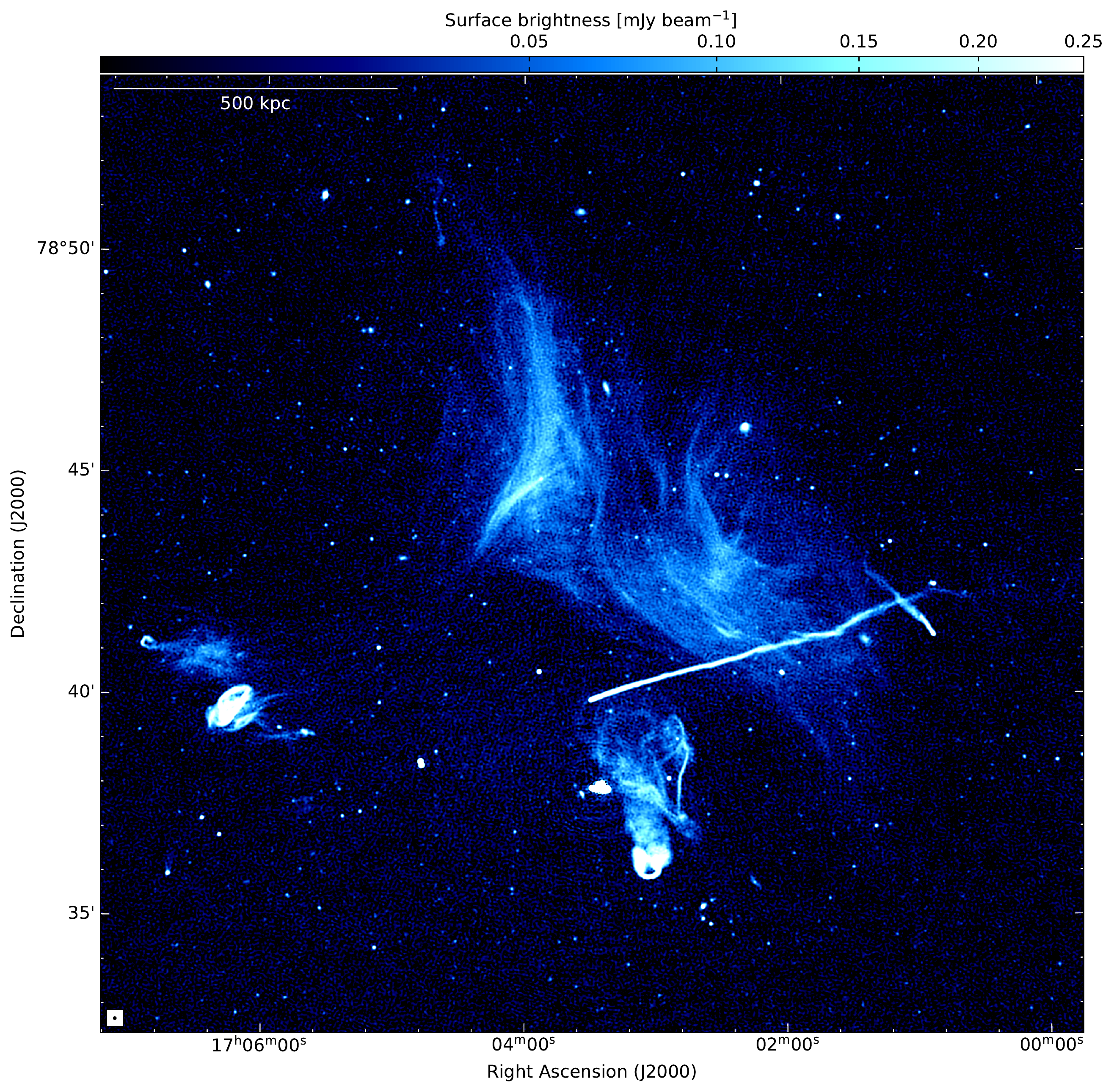}
    \vspace{-0.5cm}
 \caption{High resolution uGMRT (550-950\,MHz) image of the relic in Abell 2256. This image shows the spectacular filamentary relic and several complex radio galaxies. The image is created with {\tt Briggs} weighting using {\tt robust}=-0.6. The beam size is indicated in the bottom left corner of the image. The image properties are given in Table \,\ref{imaging}, IM4.}
      \label{high_res}
  \end{figure*}   

  \begin{figure*}[!thbp]
    \centering
    \includegraphics[width=1.00\textwidth]{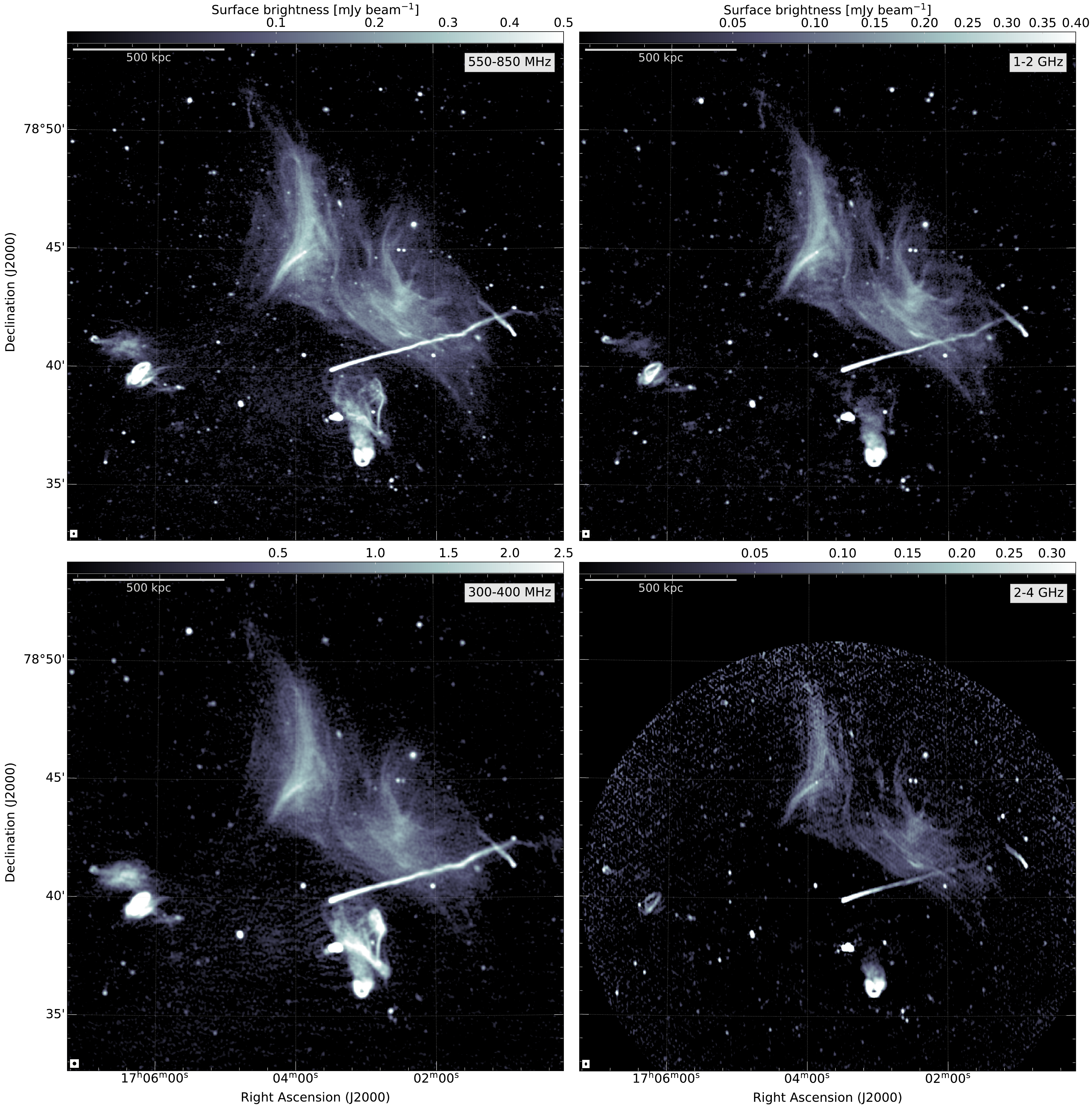}
 \caption{High resolution images of the Abell 2256 field, showing remarkable filamentary relic emission from 300\,MHz to 3\,GHz ($5\arcsec$, top left; $5\arcsec$, top right; $8\arcsec$, bottom left; $6\arcsec$, bottom right). The known large filamentary relic emission and other complex radio galaxies in the field are recovered in our new uGMRT observations. The color map uses square root scaling to enhance faint radio surface brightness structures. The beam size is indicated in the bottom left corner of each image. The image properties are given in Table \,\ref{imaging}, IM5, IM12, IM9, and IM16}
      \label{fig1}
  \end{figure*}

  \begin{figure*}[!thbp]
    \centering
    \includegraphics[width=1\textwidth]{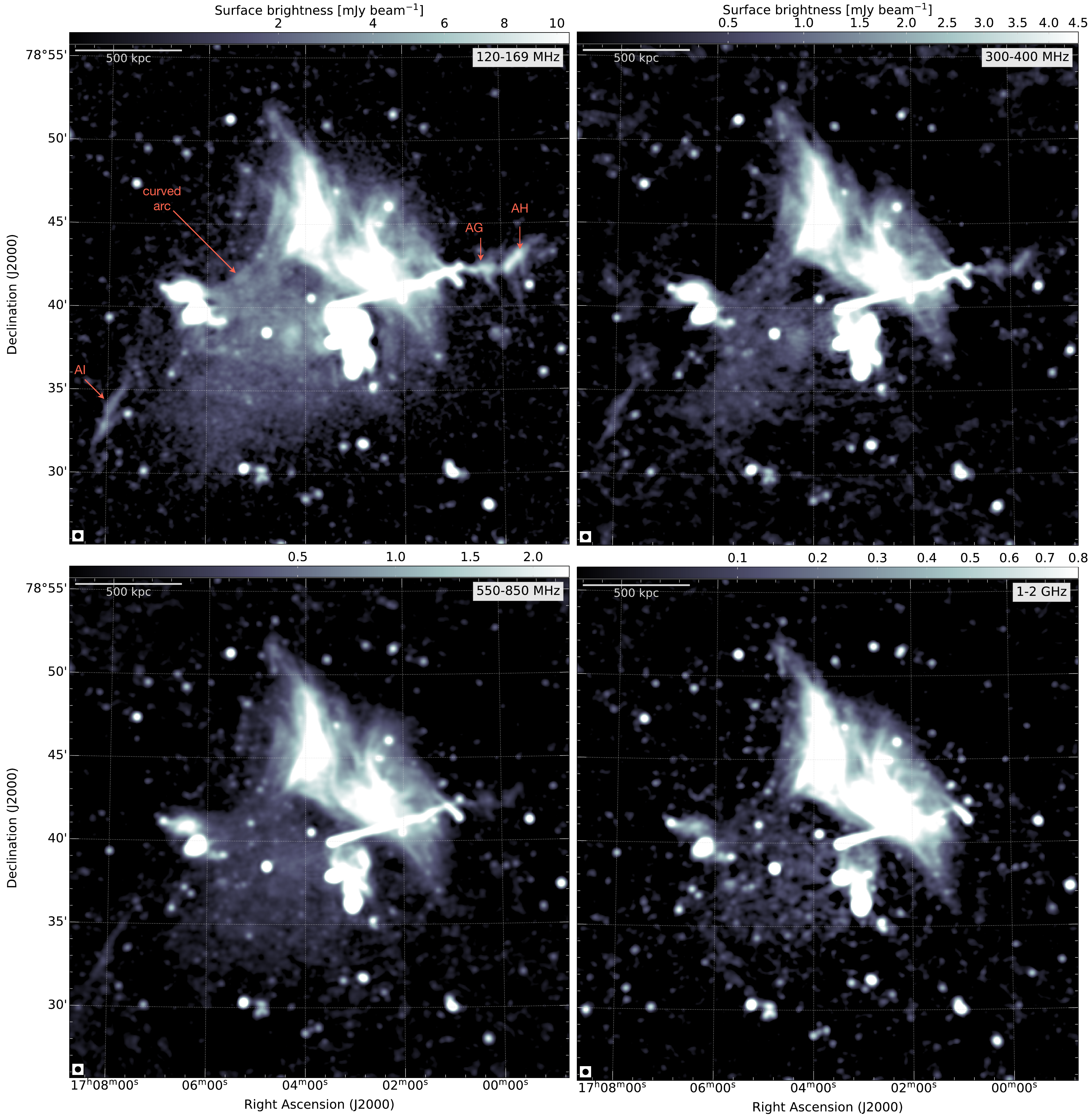}
 \caption{Low resolution ($20\arcsec$) LOFAR 144\,MHz (top-left; Osinga et al. in prep), uGMRT Band\,3 (top-right; 300-400\,MHz), uGMRT Band\,4 (bottom-left; 550-850\,MHz), and VLA L-band (bottom-right; 1-2\,GHz) images of Abell 2256 in square root scale, showing the central halo emission, AI, AG, AH, and the curved arc-shaped filament to the east. The beam size is indicated in the bottom left corner of each image. The image properties are given in Table \,\ref{imaging}, IM3, IM11, IM8, and IM15.}
      \label{low_res}
  \end{figure*}

\section{Results: continuum images}
\label{results}
Our deep $3\arcsec\times3\arcsec$ high-resolution uGMRT Band\,4 (550-850\,MHz) image of the cluster is shown in Fig.\,\ref{high_res}.  The image is created with Briggs weighting and ${\tt robust}= -0.6$ to resolve fine filaments reported at high frequency \citep{Owen2014}.  A high resolution Band\,3 image (300-400\,MHz)  is shown in Fig.\,\ref{fig1} (bottom left panel). Most of the filaments are also detected at Band\,3. In addition, the new sensitive uGMRT observations show considerable low surface brightness emission spread throughout the relic (Fig.\,\ref{fig1} left panels).   To facilitate the discussion,  we label sources in Fig.\,\ref{overlay}, following \cite{vanWeeren2009a} and \cite{Owen2014}. 

Our new observations provide the first  arcsecond-resolution (i.e., 3\arcsec-6\arcsec) image of the cluster at frequencies below 1\,GHz. Thanks to the wideband receivers, we reached an unprecedented noise level of $4 \, \rm \mu Jy\,beam^{-1}$ at a central frequency of 675\,MHz, , see Table\,\ref{imaging} for imaging parameters. To our knowledge, this is the lowest noise level obtained so far with the uGMRT in Band\,4. At Band\,3 (300-400\,MHz), we achieved a noise level of $28 \rm \mu Jy\,beam^{-1}$ at $8\arcsec$ resolution.

The most prominent source in the field is the large relic that is dominated by complex filaments on various scales (Fig.\,\ref{high_res}). These filaments were first discovered at high frequency by \cite{Owen2014}. Our uGMRT images resolve them for the first time at frequencies below 1\,GHz.  The uGMRT Band\,4 (550-850\,MHz) image of the relic looks remarkably similar to the high -frequency 1-2\,GHz VLA image. The relic consists of two main bright regions, namely, R1 (northern part) and R2 (southern part), which consist of filamentary structures and regions of diffuse emission. At a resolution of $3\arcsec$, most of the individual filaments are resolved in both axes (Fig.\,\ref{high_res}). The width of the twisted ``long filament" connecting the R1 and R2 regions, varies from $5\arcsec$ to $9.8 \arcsec$, corresponding to a physical (projected) scale of  5.7\,kpc and 11.1\,kpc, respectively. We detect at least 44 compact faint unrelated sources embedded within the relic at 675\,MHz.

To compare the morphology of the relic at sufficiently high resolution, in Fig.\,\ref{fig1}, we show the radio maps at different frequencies, namely 350\,MHz ($8\arcsec$ resolution), 675\,MHz ($5\arcsec$ resolution), 1.5\,GHz ($5\arcsec$ resolution), and 3\,GHz ($6\arcsec$ resolution). These images were created using Briggs weighting and ${\tt robust}= 0$. Since the rms noise level decreases with high values of the ${\tt robust}$ parameter, this allows us to also detect low surface brightness emission. At the $3\sigma$ level, we measure a largest linear size (LLS) of 0.7\,Mpc, 1\,Mpc, and 1.1\,kpc at 3\,GHz, 1.5\,GHz, and 675\,MHz, respectively. The entire relic covers an area of $11.1\arcmin\times6.5\arcmin$,  $14.7\arcmin\times8.8\arcmin$,  and $16.7\arcmin\times9.1\arcmin$ at 3\,GHz, 1.5\,GHz, and 675\,MHz, respectively. Therefore, a large extension of the emission is detected at lower frequencies, typical for radio relics, where the emission generally steepens with distance from the (re-)acceleration regions.

To emphasize the radio emission on a different spatial scale, we also created images at moderate resolution ($10\arcsec$) using uv-tappering; see Table\,\ref{imaging} for imaging parameters.  The resulting images at 144 MHz, 325 MHz, 675 MHz and 1.5 GHz are shown in  \ref{low_res}. A distinct morphological structure visible in these images is a low surface brightness  arc-shaped structure to the east of the cluster, first reported by \cite{Owen2014} at 1-2\,GHz (see \,\ref{low_res}). The emission is mainly visible in low resolution images. No  feature is detected in the {\it Chandra} and {\it XMM-Newton} X-ray images \citep{Ge2020} at that location. The arc-shaped feature is apparently connected to the large relic, the central halo emission, and the source F. Based only on the visual appearance, whether the connection between these features is real or just in projection is not obvious. 

In addition to the large filamentary relic, the cluster is known to host a radio halo and several complex radio galaxies (B, C, F1, F2, F3, A, and I). In both Band\,3 and Band\,4 images, we recovered these sources, see Figs.\,\ref{fig1} and \ref{low_res}. To the southeast of the cluster is a linear structure, labeled as AI in Fig.\,\ref{low_res}. It is located at a projected distance of about 1~Mpc from the cluster center and was detected previously by \cite{vanWeeren2009a} and \cite{Intema2009}. At 350\,MHz, the LAS of the source is 403\arcsec corresponding to a physical size of 455\,kpc. We measure a flux density of $2.5\pm0.1\,\rm mJy$ at 675\,MHz. The source is not detected at 1-2\,GHz. The physics of the halo emission, AI, curved-arc, and individual radio galaxies will be presented in a subsequent paper (Rajpurohit et al. in prep).

\begin{figure*}[!thbp]
    \centering
    \includegraphics[width=0.477\textwidth]{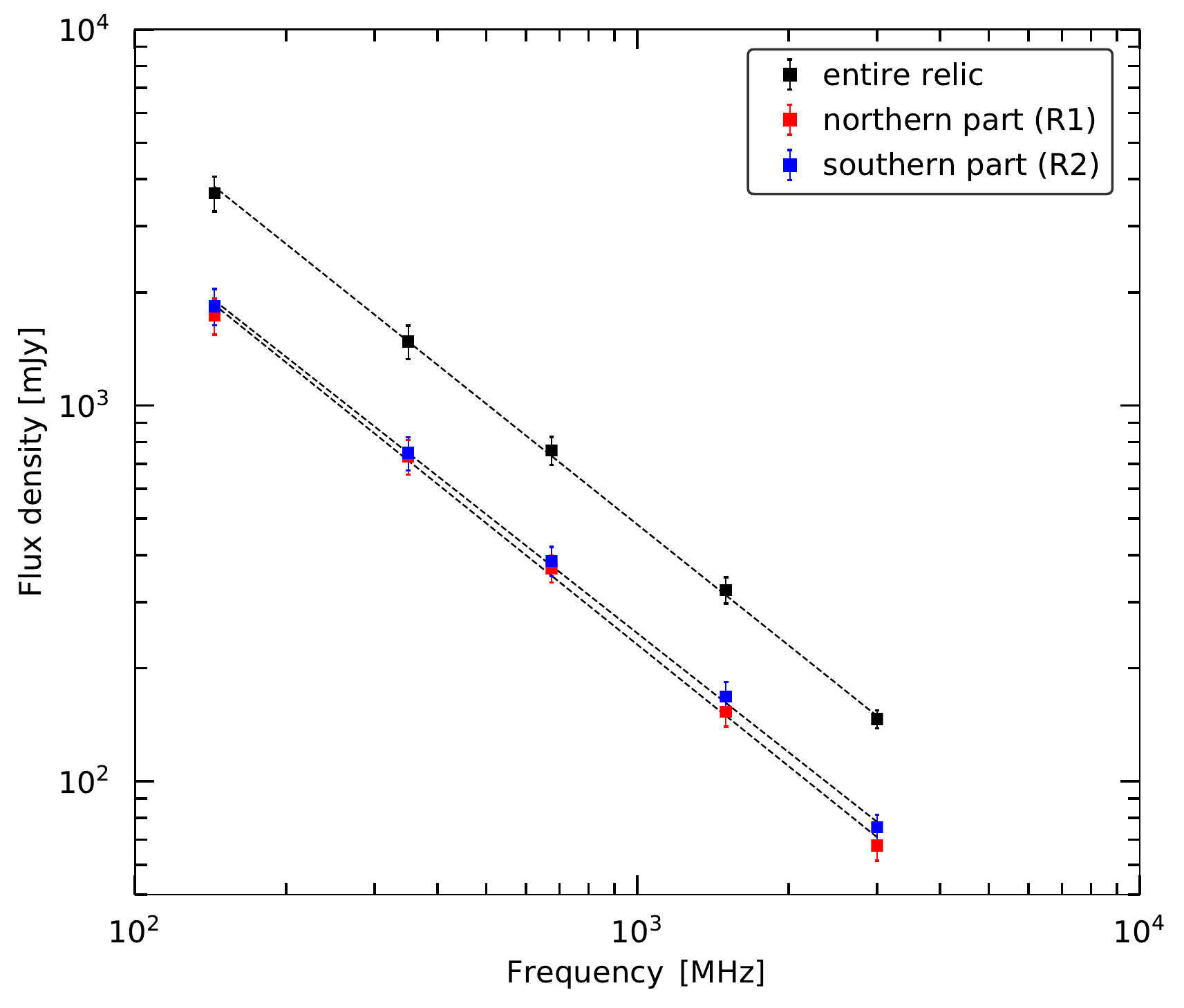}
      \includegraphics[width=0.51\textwidth]{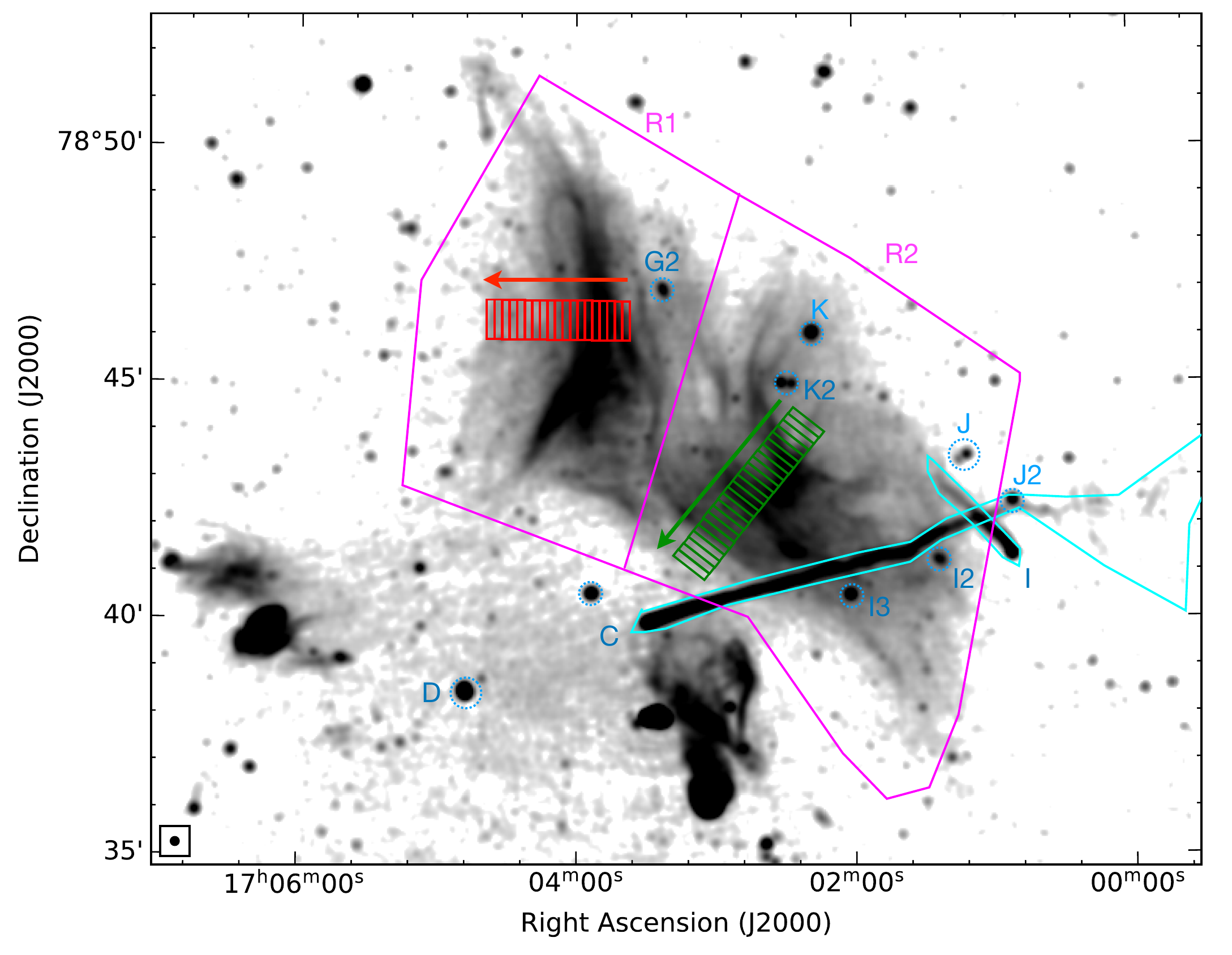}
 \caption{{\textit Left}: Integrated spectra of the relic and its subregions (R1 and R2) from 144\,MHz to 3\,GHz. The overall spectrum of the entire relic is well described by a single power-law and has a slope of $\alpha = -1.07\pm0.02$. The  subregions R1 (northern part) and R2 (southern part) also follow power-law distributions with $\alpha_{\rm R1}= -1.08\pm0.02$ and  $\alpha_{\rm R2}= -1.05\pm0.02$. \textit{Right}: Region and box distributions across the relic overlaid on the uGMRT  total intensity map at 10\arcsec resolution. The magenta regions are used for extracting the flux densities of the relic and its subregions. Flux density contributions from compact sources (shown with blue circles) and long tail angle galaxies C and I (shown with cyan) were manually subtracted from the total relic flux density. The red and green rectangular boxes were used to study the spectral index and curvature profiles shown in Fig.\,\ref{profiles}. The arrows show the direction of increasing region number for the spectral index and curvature profiles (used in Fig.\,\ref{profiles}). The width of the boxes used to extract the indices is 11.3\,kpc.}
  \label{spectra}
\end{figure*}

\setlength{\tabcolsep}{4pt}
\begin{table*}[htp]
\caption{Properties of the diffuse radio sources in the cluster Abell\,2256.}
\begin{center} 
\begin{tabular}{*{10}{c}}
\hline \hline
\multirow{1}{*}{Source} &\multirow{1}{*}{LOFAR (144\,MHz)} & \multicolumn{2}{c}{uGMRT (300-850\,MHz)} & \multicolumn{2}{c}{VLA (1-4\,GHz)  }&\multirow{1}{*}{$\rm LLS^{\dagger}$} & \multirow{1}{*}{$\alpha$}$^{\dagger\dagger}$& \multirow{1}{*}{$P_{1.5\,\rm GHz}$} \\
 \cline{3-6} 

& $S_{\rm144\,MHz}$ &$S_{\rm350\,MHz}$&$S_{\rm675\,MHz}$&$S_{\rm1.5\,GHz}$&$S_{\rm3\,GHz}$ &&&\\
  & (mJy) & (mJy) & (mJy)&  (mJy) & (mJy) &(Mpc) &&($10^{24}\,\rm W\,Hz^{-1}$) \\
  \cline{2-3} \cline{4-5}\cline{6-7}
  \hline 
Entire relic & $3580\pm390$ &$1490\pm159$& $760\pm65$& $323\pm26$&$145\pm8$&$\sim1.1$ &$-1.07\pm0.02$&$2.8$ \\ 
R1&$1740\pm190$ &$732\pm77$&$368\pm30$& $154\pm13$& $67\pm6$&-&$-1.08\pm0.02$&$-$\\ 
R2  &$1839\pm205$ &$748\pm79$&$386\pm35$& $169\pm16$&$76\pm6$&-&$-1.05\pm0.02$&$-$\\ 
\hline 
\end{tabular}
\end{center} 
{Notes. Flux densities were extracted from images created with ${\tt roboust}=-0.5$ and a \textit{uv}-cut. The relic flux density values were extracted from $10\arcsec$  resolution radio maps corresponding to images IM2, IM7, IM10, IM14, and IM17 (For imaging properties see Table\,\ref{imaging}. The regions where the flux densities were extracted are indicated in the right panel of Fig.\,\ref{spectra}. Absolute flux density scale uncertainties are assumed to be 10\% for LOFAR and Band3, 5\% for the uGMRT Band4, and 2.5\% for the VLA L- and S-band data. $^{\dagger}$The LLS measured at 675\,MHz; $^{\dagger\dagger}$ the integrated spectral index obtained by fitting a single power-law fit. }
\label{Tabel:Tabel2}   
\end{table*}

\section{Analysis and discussion}

\subsection{Filamentary structures}
\label{relic_analysis}

The origin of filaments in the Abell 2256 relic is still a mystery. High resolution radio observations of some other radio relics also show the presence of enigmatic kiloparsec scale filamentary structures in the form of threads, twisted ribbons, bristles. For example, in the Toothbrush \citep{Rajpurohit2018,Rajpurohit2020a}, Sausage \citep{Gennaro2018}, and MACS\,J7017+35 \citep{vanWeeren2017b} relics. However, the relic in Abell 2256 shows many long, pronounced, complex filaments stretching across the entire relic. Interestingly, similar complex but rather small-scale filaments are observed in the ``brush'' region of the Toothbrush relic \citep{Rajpurohit2018,Rajpurohit2020a}. It could be that the filaments found in the Toothbrush are large, similar to the Abell 2256 relic, but because it is seen edge-on, we only see the ends of the filaments.

It is challenging to understand the nature of filamentary structures in relics since  the geometry/topology of the relativistic plasma is affected by projection effects. The observed filaments could be substructures of a complex shock front, possibly highlighting the underlying distribution of Mach numbers \citep{Hong2015,Roh2019,Wittor2019,Paola2021} and a range of electron acceleration efficiencies by shock acceleration \citep{wi17}. Conversely, they can reflect a pattern of fluctuations in the magnetic field strength and topology at the corrugated surface of a shock. MHD turbulence can also produce sheet-like large filaments \citep[e.g.,][]{Ji2016,Vazza2018,Paola2020a}. Alternatively, the patchy, sheet-like filaments could be regions containing pre-existing populations of relativistic plasma and magnetic fields, e.g., from AGN, where the magnetic fields are stretched and advected by ICM motions, including the passage of shocks \citep{Ensslin2001}. \cite{Owen2014} proposed that the filamentary relic in Abell 2256 is produced by a large-scale current sheet, sitting at the boundary between two magnetic domains. Magnetic reconnection in such current sheets could accelerate particles to high energies \cite{Guo2016}. 

Advanced numerical modeling has recently provided new, important clues to the possible dynamics of such filaments. For example, \cite{Paola2021} have recently performed high-resolution three-dimensional MHDsimulations of merger shock waves propagating through a magnetized and turbulent ICM. They found complex synchrotron emission substructures within the simulated relic, compatible with the idea that  merger shock waves, sweeping through a realistically turbulent ICM, can naturally produce numerous complex filamentary structures in relics. The morphology of those filaments depends on the strength of the shock and the upstream turbulence;  a strong shock produces a more extended turbulent magnetic region, which in turn results in bright and elongated filamentary synchrotron emission while a low Mach number shock is found to produce less disrupted patterns. In this scenario, the complex filaments in the Abell 2256 relic can thus be produced by non-uniform distribution of Mach numbers in a shock front that is propagating through a magnetized and more turbulent ICM. In addition, the presence of significant RM variations \citep{Owen2014} across the relic favors the hypothesis that the medium is turbulent at the location of the relic.

\subsection{Integrated spectrum}
\label{int_spectrum}
The knowledge of the integrated radio spectra of relics is vital to understanding the particle acceleration at shock fronts. We perform a careful integrated spectral analysis of the relic in Abell 2256.  Thanks to the LOFAR, uGMRT and VLA L- and S-band data, we can study, for the first time, the spectral index distribution (averaged and spatially resolved) across the relic in great detail over a large frequency range.

\begin{figure}[!thbp]
\centering
\includegraphics[width=0.48\textwidth]{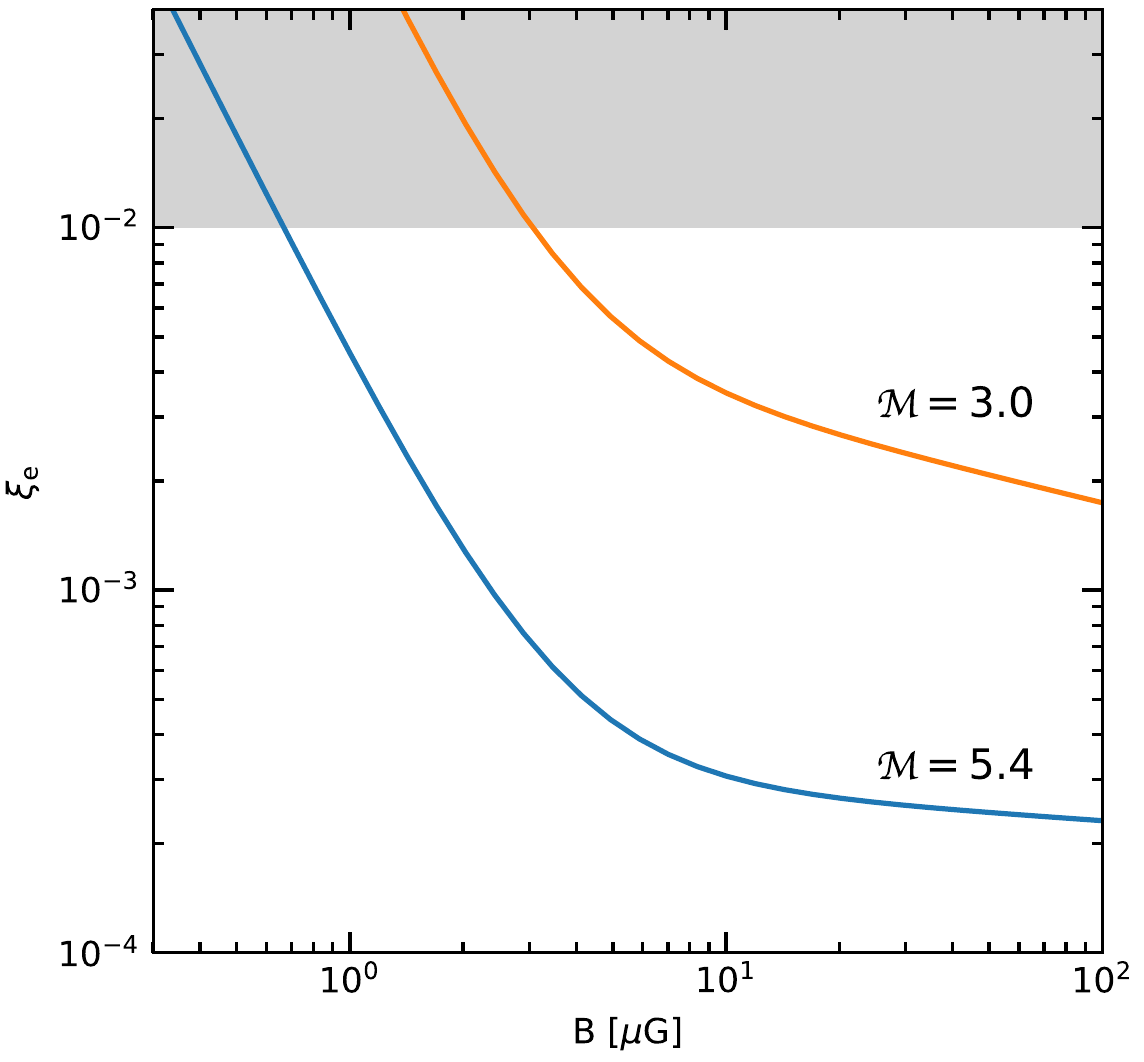}
\vspace{-0.6cm}
\caption{Energy fraction $\xi_e$ channeled into the acceleration of suprathermal electrons as a function of magnetic field. We adopted the Mach number which corresponds to the integrated spectral index, namely $\mathcal{M}=5.4$ (blue line), and a lower Mach number, $\mathcal{M}=3.0$ (orange line), which may reflect the actual shock strength, assuming that the integrated spectral index is dominated by a small fraction of the shock surface showing a higher Mach number. If only 1\,\% or less of the dissipated energy can be channeled into suprathermal electrons, the minimum required magnetic field strength is 0.7 - 3\,$\mu$G,  which is a plausible field strength for radio relics. }
\label{cc_plots2}
\end{figure}      

\begin{figure*}[!thbp]
\centering
\includegraphics[width=0.75\textwidth]{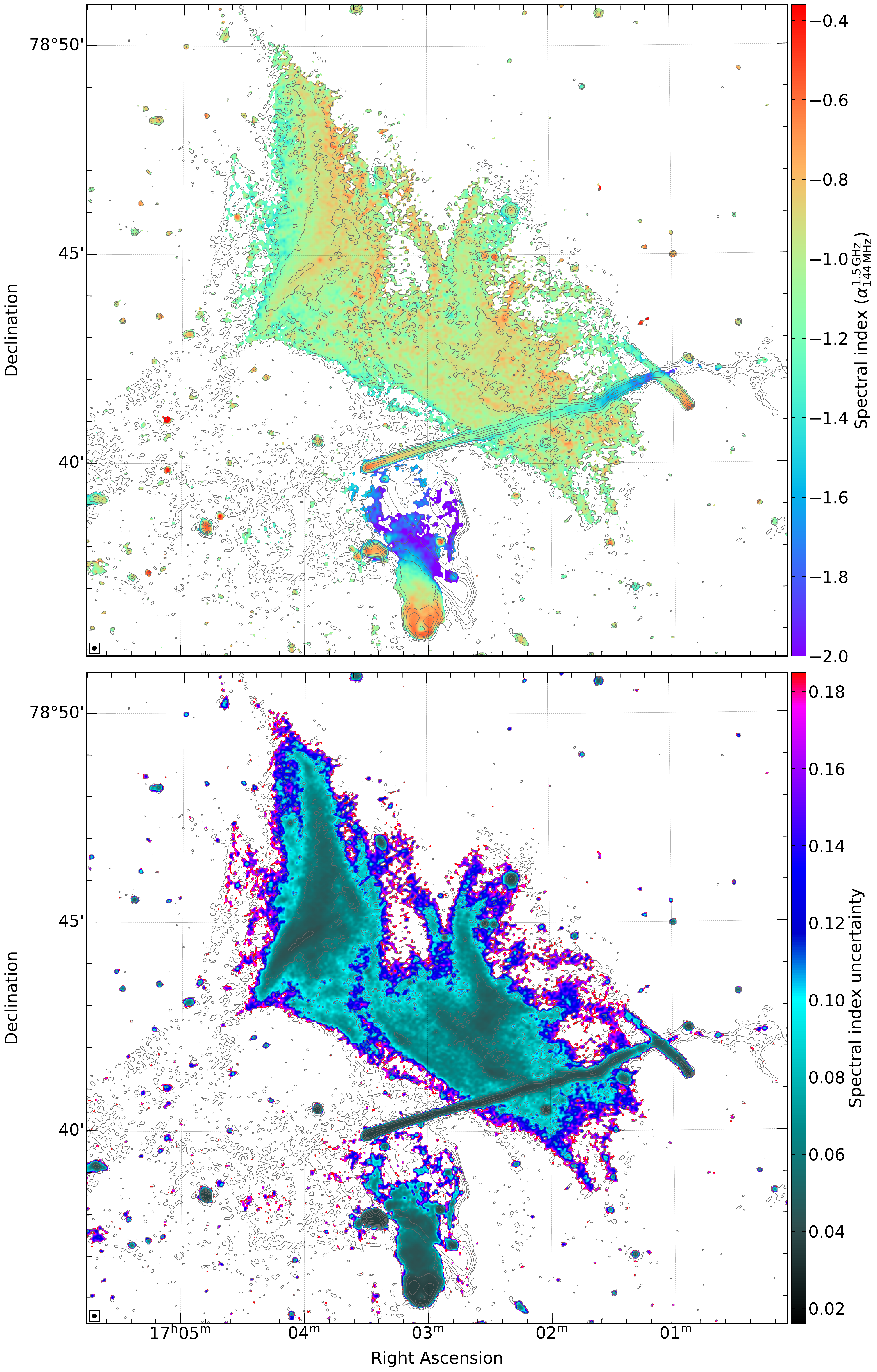}
\caption{\textit{Top:} High-resolution ($6\arcsec$) spectral index maps of the relic between 144\,MHz and 1.5\,GHz. The are spectral fluctuations across the relic. The bright filaments show spectral indices flatter than $-0.85$. Contour levels are drawn at $[1,2,4,8,\dots]\,\times\,3.0\,\sigma_{{\rm{ rms}}}$ and are from the LOFAR 144 MHz image. \textit{Bottom:} Corresponding spectral index uncertainty. The beam size is indicated in the bottom left corner of the image.}
\label{index_high}
\end{figure*}

Accurate flux density measurements of extended surface brightness sources with interferometers are challenging for several reasons: (1) missing short \textit{uv}-distances due to which the flux of the extended structures gets ``resolved out''; (2) different \textit{uv}-coverage of telescopes; (3) different weighting schemes, different resolution, and  improper deconvolution, and (4) proper subtraction of flux density contributions from unrelated sources. If not done carefully, this in turn may affect the integrated spectral index and spatially resolved spectral-index/curvature analysis. We create images with a common inner \textit{uv}-cut of $0.1\rm\,k\lambda$. Here, $0.1\rm\,k\lambda$ is the minimum well-sampled \textit{uv}-distance in the uGMRT Band\,4  data. This ensures that we are recovering the flux density on the same spatial scales at all observed frequencies.  Since different resolution and imaging weighting schemes biases the flux density measurements, we imaged each data set  at a common resolution and with ${\tt robust}=-0.5$ weighting, which ensures that we are sampling the same spatial scales of the sky. Deconvolution was always performed with multiscale cleaning  which allows to recover faint emission features properly. 

The high frequency S-band data may be affected by missing the largest angular scale by undersampling of short baselines. Therefore, we compare our flux density measurements with the values obtained with single dish observations that do not resolve-out the flux density of extended sources. We created the S-band image at 35\arcsec resolution using ${\tt robust}=0$. At 2.6\,GHz, we measure an integrated flux density of $221\pm6\,\rm mJy$ after subtracting contributions from unrelated sources (labeled in Fig.\,\ref{spectra} right panel) following \cite{Trasatti2015}. Flux density contributions from these sources were manually measured and then subtracted from the total flux density of the relic. In the same region, \cite{Trasatti2015} reported a flux density of $235\pm15\,\rm mJy$ using the Effelsberg radio telescope at 2.6\,GHz. Our S-band interferometric flux density measurement is in agreement with the value obtained from the single-dish telescope. This clearly indicates that our high frequency interferometric data are not affected by missing largest angular scale or short baselines. We note that the flux density of $221\pm6\,\rm mJy$ at 3\,GHz differ from the one reported in Table\,\ref{Tabel:Tabel2} because we used different imaging parameters and subtracted additional unrelated compact sources.

For the flux density measurements of the relic, we consider medium resolution images, namely $10\arcsec$ resolution. Since the relic is surrounded by the extended halo emission and several compact/extended sources, this resolution allows us to separate the real relic flux density from unrelated sources. While using this reasonably high resolution slightly reduces the overall integrated flux density, see \cite{Rajpurohit2018} for discussion, this effect remains the same for all data sets, thus does not affect the slope of the integrated spectral index.

The regions where the flux densities were extracted are indicated in the right panel of Fig.\,\ref{spectra}. The measured flux densities, reported in Table\,\ref{Tabel:Tabel2}, do not include the contributions from the sources C, I, K, K2, J, J2, I2, G2, and I3. Moreover, as reported in Sect.\,\ref{results}, there are 44 other very compact, faint sources embedded in the relic. For these compact sources, we measured a combined flux density of $55\rm \,mJy$, $23\rm \,mJy$, $\rm 14\,mJy$, $\rm 6.5\,mJy$, and $\rm 3\,mJy$ at 144\,MHz, 350\,MHz, 675\,MHz, 1.5\,GHz and 3\,GHz, respectively. We also subtracted the flux density contributions from those sources from the total relic flux density.

In Fig.\,\ref{spectra}, we show the resulting spectrum of the relic and its subregions. The relic follows a simple power law between 144\,MHz and 3\,GHz. The single power-law nature of the relic at frequencies below 3\,GHz is inconsistent with the earlier study by \cite{Trasatti2015}. They reported evidence for a high frequency (above 1.4\,GHz) steepening and found that the integrated spectrum can be best described by two different power laws, with $\alpha^{1.4\,\rm GHz}_{63\,\rm MHz} = -0.86\pm0.01$ and $\alpha^{10\,\rm GHz}_{1.4\,\rm GHz} = -1.02\pm0.02$. In contrast, we do not find any evidence of spectral steepening in the spectrum between 144\,MHz and 3\,GHz. In \cite{Trasatti2015}, the steepening in the flux density spectrum below 1.4\,GHz was claimed on the basis of 350\,MHz Westerbork Synthesis Radio Telescope and 63\,MHz LOFAR observations \citep{vanWeeren2012b}. Moreover, the previous integrated spectrum studies of the relic were obtained by combining single-dish observations at high frequency and interferometric observations at low frequencies, without matching the \textit{uv}-coverage. Combining such measurements can be difficult because interferometric measurements might underestimate the flux density as a result of missing short spacings, single-dish measurements lack sufficient resolution that may result in overestimation of the flux density owing to to source contamination. In addition, using non-matching \textit{uv}-coverage and/or different imaging parameters may also result in under- or over-estimation of the integrated flux density \citep[e.g.][]{Stroe2016,Rajpurohit2018}.

Our integrated spectral index analysis shows that both the whole relic and its two main subregions (R1 and R2) also follow a power-law spectrum. This is consistent with previous detailed studies of well-known bright relics over a similarly broad frequency range, principally the Toothbrush and MACS\,J0717.5+3745 relics \citep{Rajpurohit2020a,Rajpurohit2021a}.

\begin{figure*}[!htbp]
\centering
\includegraphics[width=0.8\textwidth]{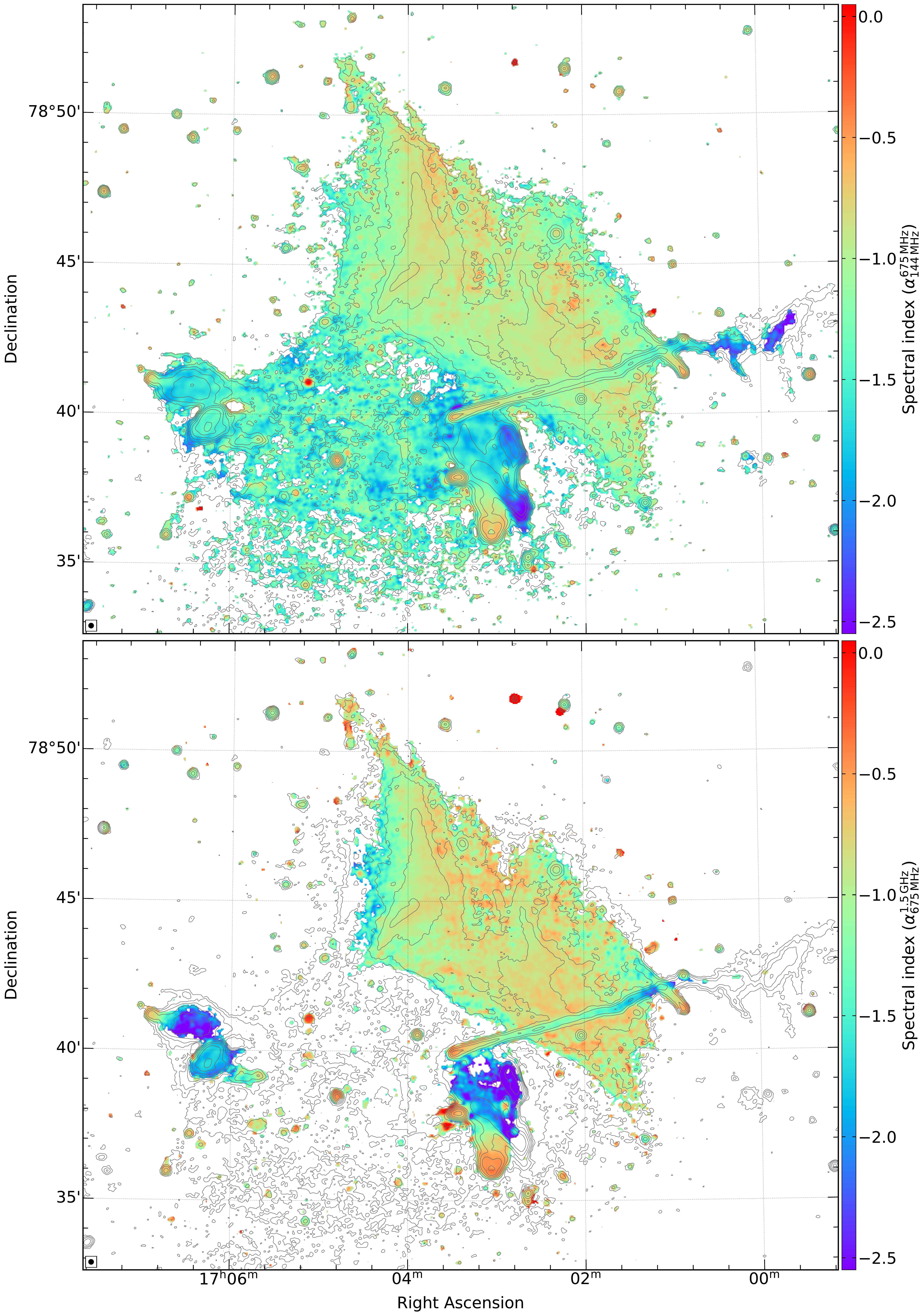}
\caption{Spectral index maps of the relic between 144\,MHz and 675\,MHz (top) and 675\,MHz and 1.5\,GHz (bottom) at 10\arcsec  resolution. A clear spectral index gradient is visible across the northern part of the relic while only a mild gradient is seen in the southern part of the relic. Contour levels are drawn at $[1,2,4,8,\dots]\,\times\,3.0\,\sigma_{{\rm{ rms}}}$ and are from the LOFAR 144 MHz image. The beam size is indicated in the bottom left corner of the each image.}
\label{index_low}
\end{figure*}

The most important result of our analysis is that the overall spectrum of the relic between 144\,MHz and 3\,GHz shows an integrated spectral index of $-1.07\pm0.02$. This differs significantly from all previously reported values where the spectral index is reported to be flatter than $-1.0$. By fitting a single power-law, \cite{vanWeeren2012b} and \cite{Trasatti2015} reported an integrated spectral index of $\alpha_{\rm 63\,MHz}^{\rm 1.4\,GHz}=-0.81\pm0.02$ and $\alpha_{\rm 63\,MHz}^{\rm 10.5\,GHz}=-0.92\pm0.02$, respectively.

If radio relics are generated by the stationary shock scenario, they are expected to exhibit a power-law spectrum with an integrated spectral index of $-1$ or steeper. An integrated spectral index flatter than $-1$ is not possible in the test-particle approach of DSA. Our new integrated spectral index value is instead consistent with the (quasi)stationary shock approximation. This result is important as it allows us to connect the Abell 2256 relic with the majority of other radio relics in the literature, and surmise that its peculiar morphology and substructure may likely be interpreted as the rare case of a bright radio relic whose shock normal is at a small angle toward the line of sight.

Using the spectral index of $-1.07\pm0.02$, we estimate the radio power of the relic. At 1.5\,GHz, the relic radio power is $P_{\rm 1.5\,GHz}=2.4\times10^{24}\rm \,W Hz^{-1}$. With this new value, the large relic in Abell 2256 fits well with the radio power versus LLS relation of other known relics.

\begin{figure*}[!thbp]
    \centering
     \includegraphics[width=.49\textwidth]{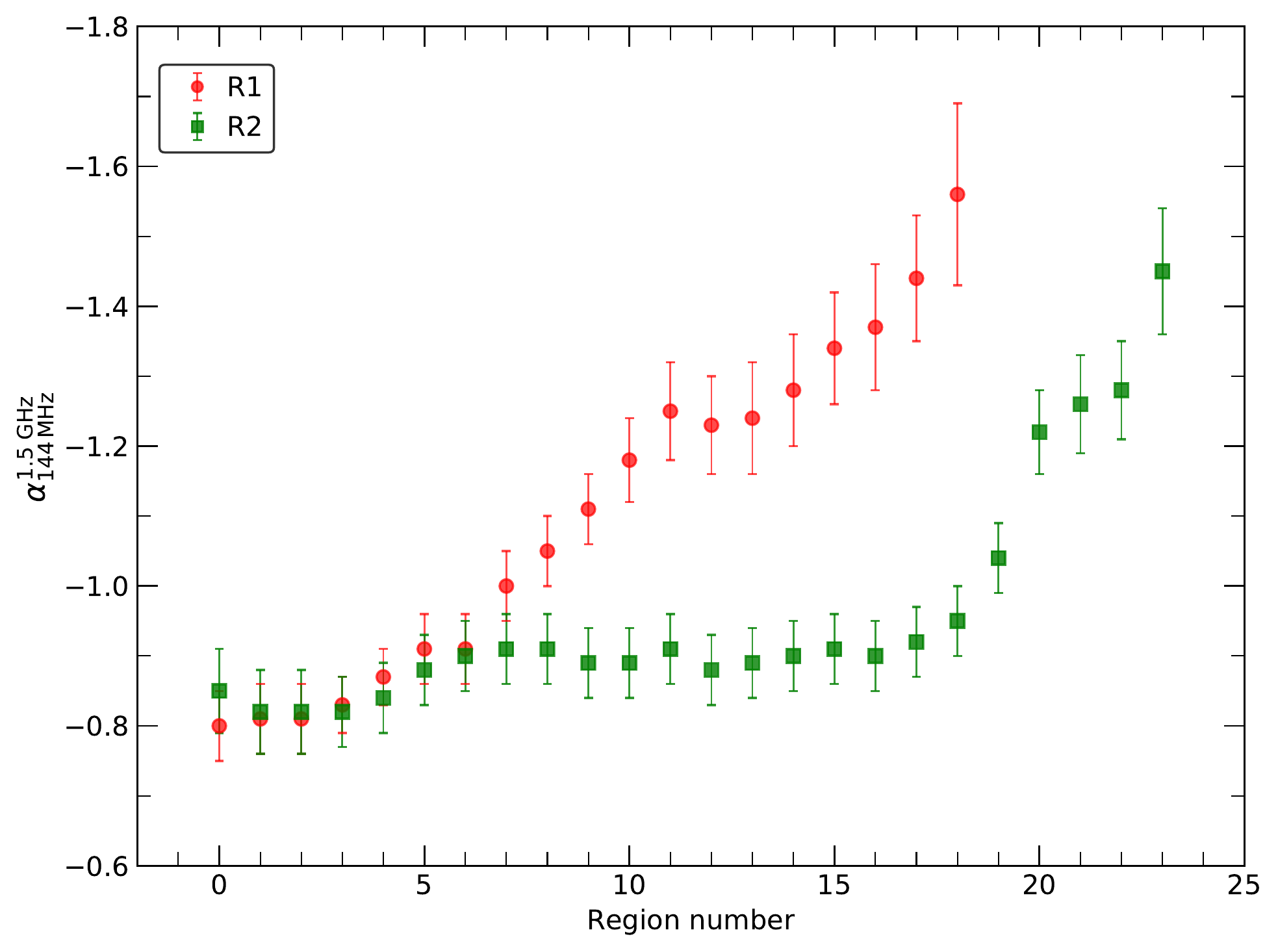}
        \includegraphics[width=.49\textwidth]{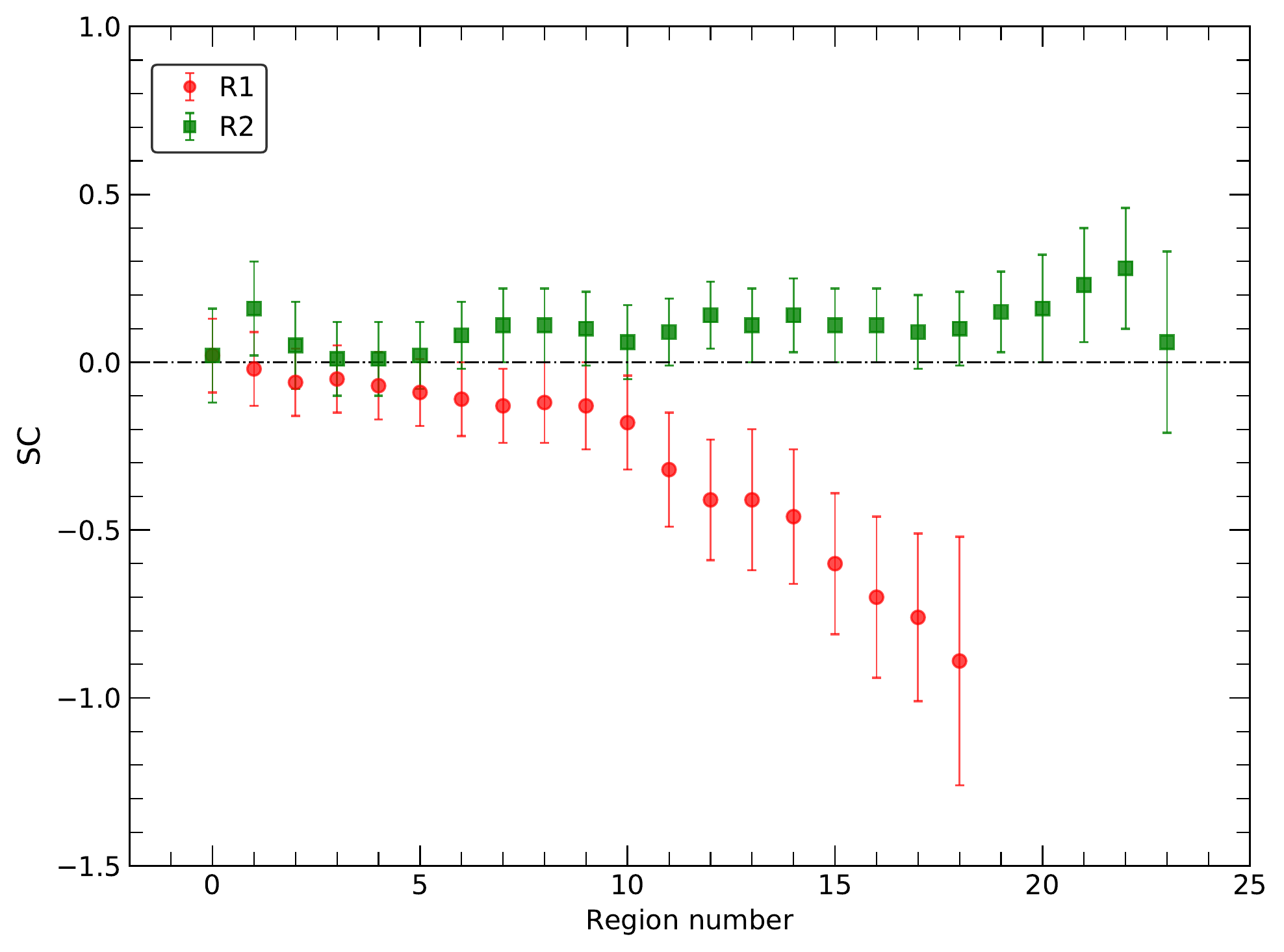}
 \caption{Spectral index (left) and curvature (right) profiles across the northern and southern parts of the relic. The spectral curvature is obtained using 144\,MHz, 675\,MHz, and 1.5\,GHz images. The R1 region of the relic shows a spectral index  and a convex curvature gradient toward the cluster center. By contrast, the R2 region shows a constant concave curvature. The total length of the red region is about 214\,kpc, while the green region is 271\,kpc. The absence of significant negative curvature across the large part of the relic provides evidence that it is very likely seen face-on. Regions used for extracting the spectral indices are shown in the right panel of Fig.\,\ref{spectra}. }
      \label{profiles}
\end{figure*}  

According to DSA in the test-particle regime, the integrated index is related to the Mach number of the shock as \citep{Blandford1987}: 
\begin{equation}
\mathcal{M}=\sqrt{\frac{\alpha_{\rm int}-1}{\alpha_{\rm int}+1}}.
\label{int_mach}
\end{equation} 
The measured integrated spectral index for the relic in Abell 2256 suggests a
shock of Mach number $\mathcal{M}=5.4^{+1.0}_{-0.6}$. If the integrated radio spectral index corresponds to the Mach number of the shock, this is one of the strongest shocks inferred in any radio relic to date. Despite the different morphology, substructures, and location, the R1 and R2 regions of the relic also show similar spectral index values, namely $\alpha_{\rm R1}= -1.08\pm0.02$ and  $\alpha_{\rm R2}= -1.05\pm0.02$. 

The Mach number may fluctuate across the merger shock front, hence, the average Mach number depends on how the Mach number distribution is averaged. The Mach numbers obtained from the integrated radio spectral indices tend to be higher than the average hydrodynamical Mach number, since the radio luminosity increases with the Mach number \citep{Wittor2021}.  Therefore,  the very high  Mach number obtained from our radio analysis for the relic in Abell 2256 may result from the fact that those patches of the shock front with a high Mach number contribute more to the emission than those with a lower Mach number.

\subsection{Energy fraction channeled into acceleration of suprathermal electrons}

The synchrotron luminosity of a radio relic depends on that energy fraction, which is channeled into the acceleration of supra-thermal electrons,  dissipated at the shock front. This energy fraction comprises the energy of all electrons which are more energetic than the electrons in the thermal pool. For a power-law energy distribution of the suprathermal electrons the synchrotron luminosity depends on the properties of the downstream medium according to \cite{Hoeft2007}:

\begin{eqnarray}  
  L_{\nu} 
  & = &  
  C \cdot   
  \xi_{\rm e} \cdot
  \frac{A}{\rm Mpc^2} \cdot 
  \frac{n_{\rm e,\,d}}{\rm 10^{-4} \,cm^{-3}} \cdot 
  \left( \frac{T_{\rm d}}{\rm 7\,keV} \right)^{\frac{3}{2}}\,\cdot\,\Psi({\cal M},T_{\rm d})
  \nonumber 
  \\
  && \quad 
  \cdot 
  \left( \frac{\nu}{1.4 \, \rm GHz} \right)^{\alpha} \cdot
  \left( \frac{B}{\rm \mu G} \right)^{-1 - {\alpha}} \cdot
  \left( \frac{B_{\rm CMB}^2}{B^2} + 1 \right)^{-1} 
  \label{eff}
\end{eqnarray} 
where $\xi_{\rm e}$ is the fraction of the kinetic energy dissipated at the shock front channeled into the acceleration of suprathermal electrons, $A$ is the surface area of the relic, $n_{\rm e,\,d}$ is the downstream electron density, $T_{\rm d}$ is the downstream electron temperature, $B$ is the magnetic field strength, $B_{\rm CMB}$ is the field strength equivalent to the cosmic microwave background energy density, $ B_{\rm CMB} =  3.24 \,(1+z)^{2}\,\mu G$, and $\Psi({\cal M}, T_{\rm d})$ comprises all Mach number dependencies. For the relic, we adopt  $n_{\rm e,\,d}=1\times10^{-3}\,\rm cm^{-3}$ and $T_{\rm d}=8\,\rm keV$ \citep{Trasatti2015,Ge2020}. The other measured values are $A=0.8\,\rm Mpc^2$, $\alpha=-1.07$, and $\Psi({\cal M})=0.76$ (assuming a shock of $\mathcal{M}=5.4$). The constant C is $1.28\times10^{27}\,\rm W \, Hz^{-1}$ \citep{Rajpurohit2021c}.

Both the magnetic field strength and the energy fraction $\xi$ are little constrained. We follow \cite{Botteon2020} and plot in Fig.\,\ref{cc_plots2} the fraction $\xi_{\rm e}$ which is necessary to obtain a radio luminosity consistent with the observation as a function of magnetic field strength. As discussed above, \citet{Wittor2021} showed that the Mach number derived from the integrated spectral index might be higher than the actual average hydrodynamical shock strength. Therefore, we also show the energy fraction $\xi_{\rm e}$ assuming a Mach number of 3.0, in accord with the typical volume averaged Mach number in \citet{Wittor2021}. We keep all other parameters unchanged. For a magnetic field strength of a few $\mu$G or larger in the radio relic region, a energy fraction of 1\,\% or less needs to be channeled into the acceleration of supra-thermal electrons, including the relativistic electrons giving rise for the observed synchrotron emission. 

In the framework of DSA, about a few percent of the energy dissipated at the shock front needs to be transferred to the supra-thermal electrons accelerated at the shock front to explain the radio power of several relics, for example the Toothbrush, the Sausage, and the relic in MACS\,J0717.5+3745 \citep{Botteon2020,Rajpurohit2020a}. It has thus been suggested that the high acceleration efficiency required to match the observed radio powers requires pre-existing population of fossil electrons \citep{Markevitch2005,Pinzke2013,Kang2016a,Vazza2015,Botteon2020}. Recent works by \cite{Kang2021} and \cite{Inchingolo2021} suggest that the electron re-acceleration by multiple shocks by DSA could also enhance the acceleration efficiency. The large relic in Abell 2256 is among the few bright relics, which can be explained by the standard relic formation scenario. 

\begin{figure*}[!thbp]
    \centering
    \includegraphics[width=0.8\textwidth]{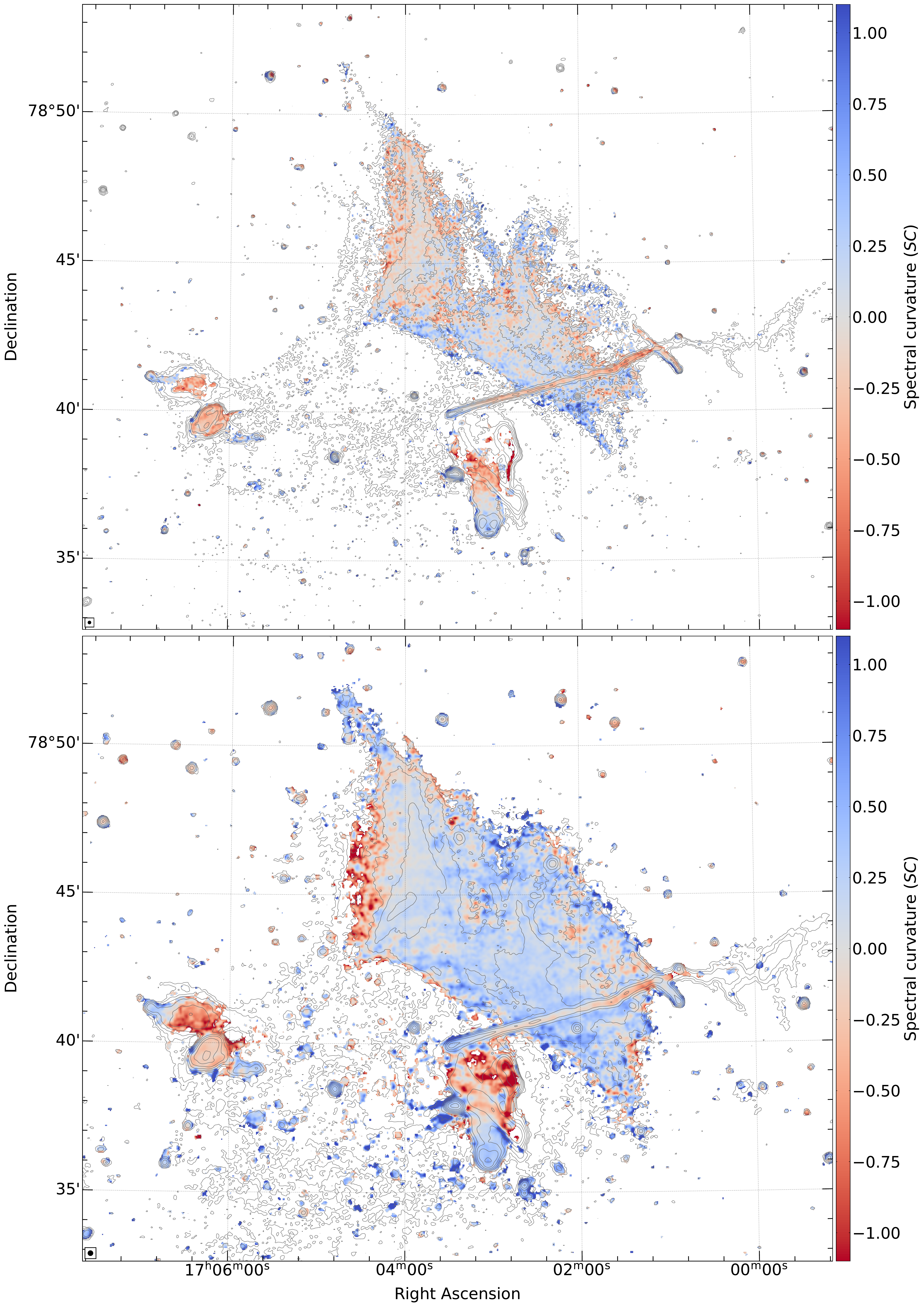}
 \caption{Three frequency spectral curvature ($SC$) maps of the relic at  6\arcsec (top) and 10\arcsec (bottom) resolutions. The relic shows both negative and positive curvature. The curvature is mostly close to zero and positive for the southern part of the relic (R2) while the negative curvature is visible at the northern part (R1). The beam size is indicated in the bottom left corner of the each image.}
      \label{curvature_map}
\end{figure*}

\subsection{Spatial spectral index distribution}
\label{spectral_index_maps}
The large size and high surface brightness allow us to conduct the most detailed spatially resolved spectral analysis of the relic in Abell 2256 to date. We create a spectral index map at the highest possible common resolution (i.e., $6\arcsec$ resolution, corresponding to a physical scale of 7\,kpc) using 144\,MHz LOFAR and 1.5\,GHz VLA data. The relic shows complex large-scale filaments, therefore this resolution also allows us to minimize the mixing of emission with different spectral properties within a single radio beam. Moreover, as the relic shows emission on different spatial scales, we also create spectral index maps at a medium resolution of $10\arcsec$ (corresponding to a physical scale of 11\,kpc). The same maps are used to create the curvature maps and radio color-color plots presented in Sect.\,\ref{curvature_anaylsis}. We create $10\arcsec$ spectral index maps at two frequency sets: between 144\,MHz and 675\,MHz and between 675\,MHz and 1.5\,GHz. For all spectral index maps, we considered only pixels with a flux density above $3\sigma_{\rm rms}$ in each map.

The resulting high resolution spectral index map between 144\,MHz and 1.5\,GHz is shown in Fig.\,\ref{index_high}.  The relic is indeed made up of different values of the spectral index, varying from about $-0.6$ to $-1.6$. There are also small-scale fluctuations in the spectral indices, at least at the order of the beam size, i.e., 7\,kpc, in particular in the R2 region. Such fluctuations in the spectral index distribution are also seen in simulations when the medium is more turbulent \citep{Paola2021}. Moreover, it seems that the filaments have different spectral index values with respect to the low surface brightness regions of the relic. 

The textbook examples of  radio relics, such as the Sausage and the Toothbrush, show clear spectral index gradients in downstream areas, reflecting the aging of the relativistic electron population, while the shock front propagates outward \citep{vanWeeren2010,vanWeeren2012a,Stroe2013,Rajpurohit2020a,Gennaro2018,deGasperin2020}. Even at such a high resolution, we do not find any clear uniform trends of the spectral index gradient across the entire relic in Abell 2256. However, the northern part of the relic apparently shows a hint of a spectral index gradient from west to east. The main difference with respect to the Sausage and the Toothbrush relics, is that the the relic in Abell 2256 is believed to be seen nearly face-on, while the other two are close to edge-on.

The medium-resolution spectral index maps for a set of two pairs of frequencies are shown in Fig.\,\ref{index_low}. In these maps, we see more emission, for example, at the eastern end of R1, which is resolved-out in the high resolution spectral index map shown in Fig.\,\ref{index_high}. The steepest spectral indices are seen at the eastern end of the R1 region of the relic, namely about $-2$. The spectral index gradient in R1 is clearly visible in these maps. The similar gradient is not observed in the R2 region of the relic.

In Fig.\,\ref{profiles} (left panel), we show the spectral index profiles extracted between 144\,MHz and 1.5\,GHz for the northern and southern part of the relic, averaged over rectangular boxes of width 11\,kpc. These boxes are indicated in the right panel of Fig.\,\ref{spectra}. The spectral index across the southern part of the relic is relatively constant except for the last five data points, which may be contaminated by the emission from source B and the halo. In contrast, a clear spectral index gradient is evident in the northern part. This part shows a spectral index steepening up to $-1.5$ over a distance of about 214\,kpc. The difference in the spectral index trends between the northern and southern parts of the relic could be due to projection effects.

The spatially resolved spectral index maps are often used to measure the injection index at the shock front. Since the large part of the relic in Abell 2256 is seen close to face-on, it is very difficult to measure the spectral index variations or curvature. In such cases, the spectral index and curvature are highly affected by the superposition of the relic emission along the line of sight and, and thus do not allow to  one to discriminate the shock edge from the downstream regions.

\subsection{Spectral curvature}
\label{curvature_anaylsis}
There are only a handful of relics where the spectral curvature distribution is reported. The difficulty in studying spectral curvature is that, in general, any curvature in the spectrum is very gradual and therefore requires sensitive and high-fidelity observations over a large frequency range at a matched spatial resolution. The sensitive LOFAR, uGMRT Band\,4 and VLA L-band allow one to perform a high spatial resolution curvature analysis of the relic. We use the spectral index maps created in Sect.\,\ref{spectral_index_maps}. The pixel-wise curvature maps are created as:   
\begin{equation}
SC = -\alpha_{\rm 144\,MHz}^{\rm 675\,MHz}+\alpha_{\rm 675\,MHz}^{\rm 1.5\,GHz}
\end{equation}
According to this convention, the curvature is negative for the convex spectrum and positive for the concave spectrum. The resulting curvature maps at two different resolutions ($6\arcsec$ and $10\arcsec$ resolutions) are  shown in Fig.\,\ref{curvature_map}.  The curvature distribution is quite complex in the relic and varies mainly from $-1.0$ to $1.0$.

As visible in the high resolution curvature map, the relic shows both convex and concave curvatures, see the top panel of Fig.\,\ref{curvature_map}. There are also small-scale curvature variations along the length of the relic with a size of the order of the beam size, i.e. 7\,kpc. In the medium resolution  curvature map (bottom panel of Fig.\,\ref{curvature_map}), the northern part of the relic shows a clear convex curvature gradient: the curvature increases from east to west, reaching to values of $-1.0$. The eastern end of the R1 region shows the maximum convex curvature.
 
The curvature profiles for the northern and southern part of the relic, as measured in the same boxes as the spectral index profiles (see Fig.\,\ref{spectra}), are shown in the right panel of Fig.\,\ref{profiles}. As seen in the spatially resolved curvature maps, the northern part of the relic reveals a clear convex curvature gradient. In contrast, the curvature is almost constant and not significant across the southern part of the relic.

Both the shock acceleration and re-acceleration models predict an increasing curvature in the downstream regions of the relic due to the aging of electrons. To our knowledge, curvature maps are currently available for only three relics: the Sausage relic \citep{Stroe2013}, and the relics in MACS\,J0717.5+3745 \citep{Rajpurohit2021b} and Abell 2744 \citep{Rajpurohit2021c}. These relics are believed to be seen close to edge-on. In contrast to the relic in Abell 2256, a clear curvature gradient is seen across these relics, suggesting active acceleration at the shock front and radiation losses in the downstream regions. The absence of any significant uniform curvature gradient for the entire relic in Abell 2256, therefore, strongly suggests that the southern part of the relic, the R2 region, is seen close to face-on. The R1 region is instead seen edge-on. 

\begin{figure*}[!thbp]
\centering
\includegraphics[width=1\textwidth]{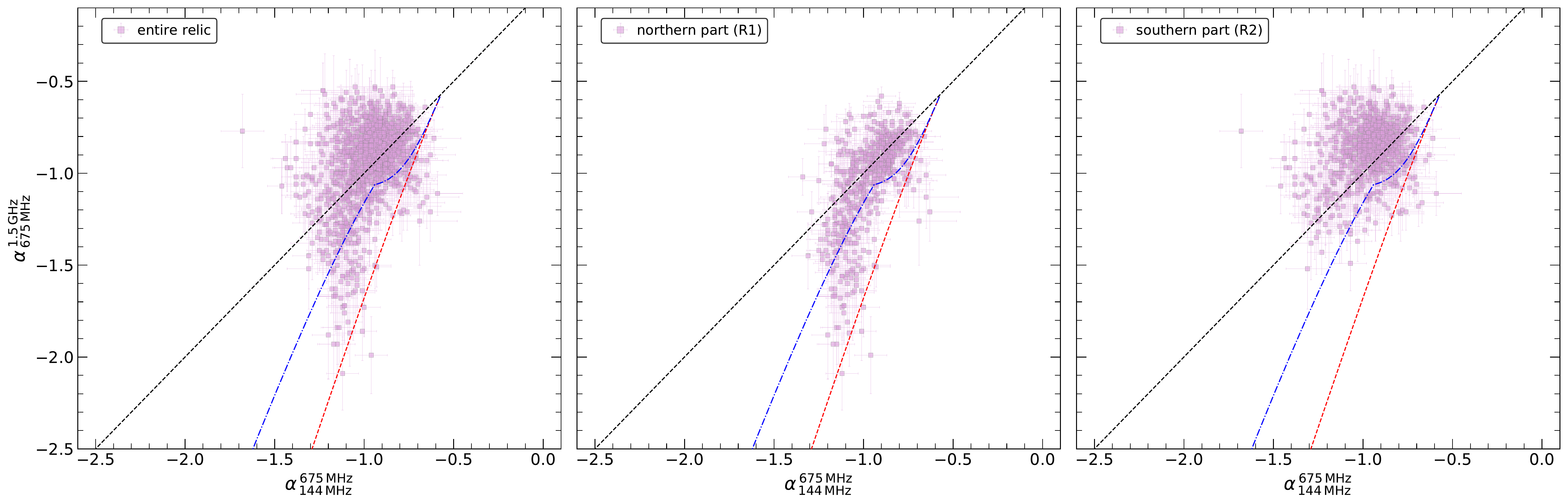}
 \caption{Radio color-color diagram of the entire relic (left) and the northern (middle) and the southern (right) parts of the relic in Abell 2256, superimposed with the JP (red dash line) and KGJP (blue dashed-dotted line) spectral aging models adopting $\alpha_{\rm inj}=-0.57$. To extract spectral index values, we create square boxes with a width of $10\arcsec$ corresponding to a physical size of about 11\,kpc. Compared to the other known relics, the Abell 2256 relic shows a complicated curvature distribution. Most of the points are clustered around the power-law or above it, implying that the large part of the relic is very likely seen face-on.}
\label{cc_plots1}
\end{figure*}   

\subsection{Radio Color-Color diagrams}
\label{cc_anaylsis}
To investigate the spectral shape of the relic, we  make use of radio color-color diagrams \citep{Katz1993}. These diagrams represent the relation between low and high frequency spectral indices for various positions in the source. Radio color-color diagrams provide important information about the electron and magnetic field distribution, mixing of different emission features, and the role of projection effects \citep[e.g.,][]{Rajpurohit2020a,Rajpurohit2021a}. We use $10\arcsec$ resolution 144\,MHz, 675\,MHz and 1.5\,GHz radio maps to construct the color-color diagrams. The entire relic is covered by a grid of square shaped boxes of width $10\arcsec$, i.e., similar to the beam size. We only included regions where the radio source brightness is above $3\sigma_{\rm rms}$ at all three frequencies. The resulting plots are shown in Fig.\,\ref{cc_plots1}.  Each data point in these plots comes from the integration of the flux within each box. 

As found in the spatially resolved curvature maps, the relic shows a complex spectral shape. Unlike the integrated spectra (obtained by averaging over a large volume of relic emission), the individual data points deviate from a simple power law (dashed lines in Fig.\,\ref{cc_plots1} where $\alpha_{144\,\rm MHz}^{675\,\rm \rm MHz}=\alpha_{675\,\rm MHz}^{1.5\,\rm GHz}$), indicating the presence of clear curvature. It is clear that most of the data points are clustered around and above the power-law line, suggesting power-law and concave spectra rather than the typical convex spectra found in other radio relics \citep{vanWeeren2012a,Stroe2013,Gennaro2018,Rajpurohit2020a,Rajpurohit2021a,Rajpurohit2021c}. 

Data points around the power-law line imply that those regions very likely have experienced recent particle injection. Concave behavior in the color-color plane is expected if a source has an inverted spectrum or if different particle populations are superimposed along the line of sight. As seen in the total power images, the relic is composed of a network of filaments surrounded by low surface brightness diffuse emission. Therefore, it is plausible that the concave spectra are mainly due to the complex superposition of different filamentary structures and low surface brightness emission along the line of sight.  We discuss this further in Sect.\,\ref{tomographysection} and Sect.\,\ref{SB_vs_index}.

A significant number of data points also lie far below the power-law line. Most of these points belong to the R1 region of the relic, indicating convex spectra at those locations, see Fig.\,\ref{cc_plots1} (middle panel). The steep spectrum regions, as visible in Fig.\,\ref{index_high}, show mostly convex spectra. This hints that this part of the relic is very likely seen close to edge-on, similar to the Sausage and Toothbrush relics.

\begin{figure*}[!thbp]
    \centering
    \includegraphics[width=0.99\textwidth]{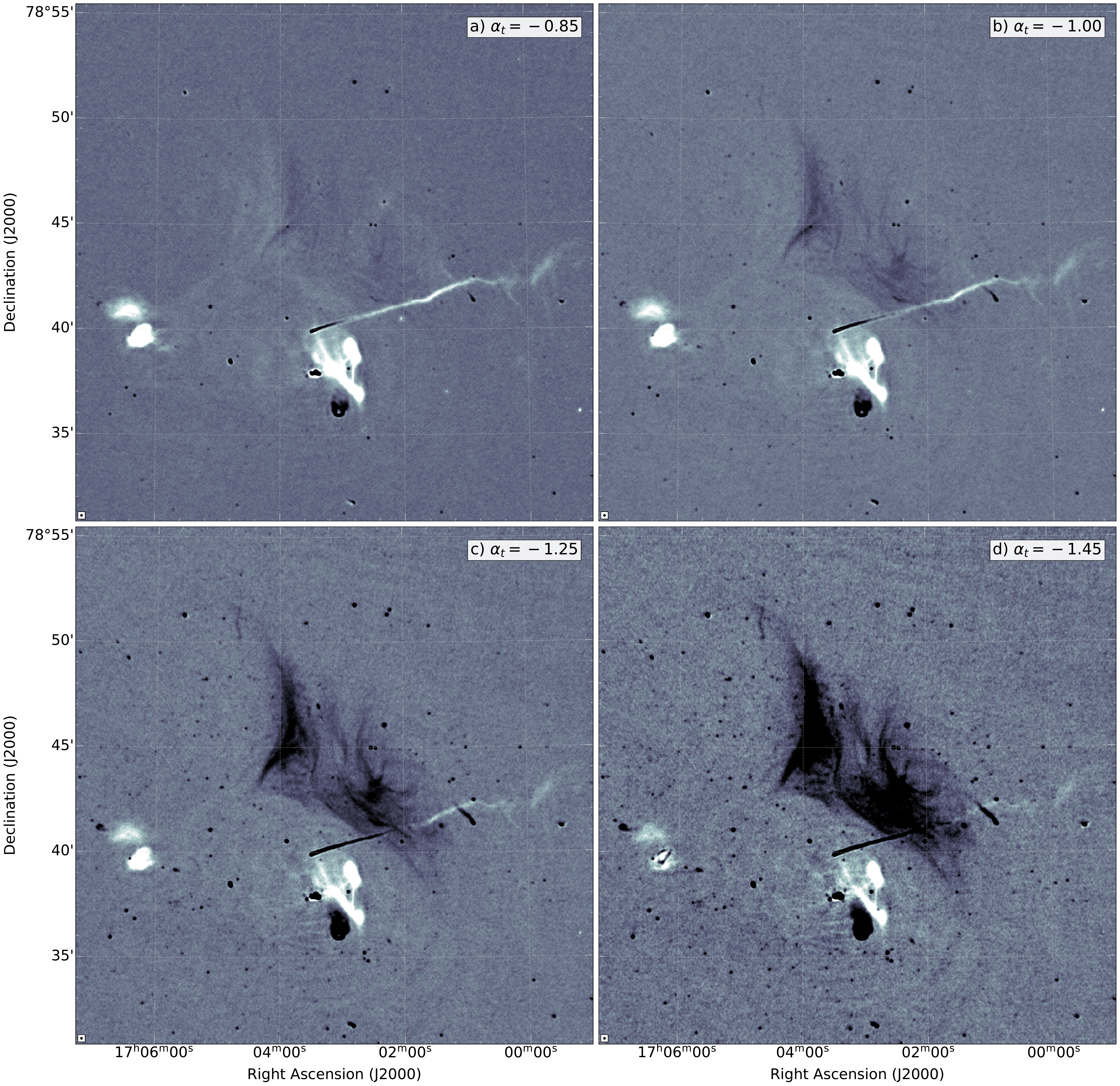} 
 \caption{Gallery of spectral tomography maps between 144\,MHz and 1.5\,GHz at $6\arcsec$ resolution. The range of $\alpha_{t}$ is $-0.85$, $-1.00$, $-1.25$, and $-1.45$. The regions with a spectrum steeper than $\alpha_{t}$ appear positive (light regions), while regions with flatter spectrum  appear negative (dark regions). These images demonstrate that there are several filamentary structures with different spectral indices. This also implies that the relic is composed of multiple overlapping structures.}
      \label{tomography}
  \end{figure*}   

The overall spectral shape in the color-color diagram for the relic in Abell 2256 is very different from the other known large relics, for example, the Toothbrush \citep{vanWeeren2012a,Rajpurohit2020a}, Sausage \citep{Stroe2013,Gennaro2018}, MACS\,J0717.5+3745 \citep{Rajpurohit2021a} and Abell 2744 \citep{Rajpurohit2021c} relics. These relics exhibit a single, continuous spectral shape throughout with a clear convex curvature gradient, indicating the acceleration of the relativistic electrons at the leading edge of the relic, where the shock front is expected, followed by radiative cooling. In contrast, the spectral shape of the Abell 2256 relic is complex and shows a broad range of spectra. Excluding a part of the R1 region, the data points are mostly clustered around and above the power-law line.

\begin{figure*}[!thbp]
\centering
\includegraphics[width=1.0\textwidth]{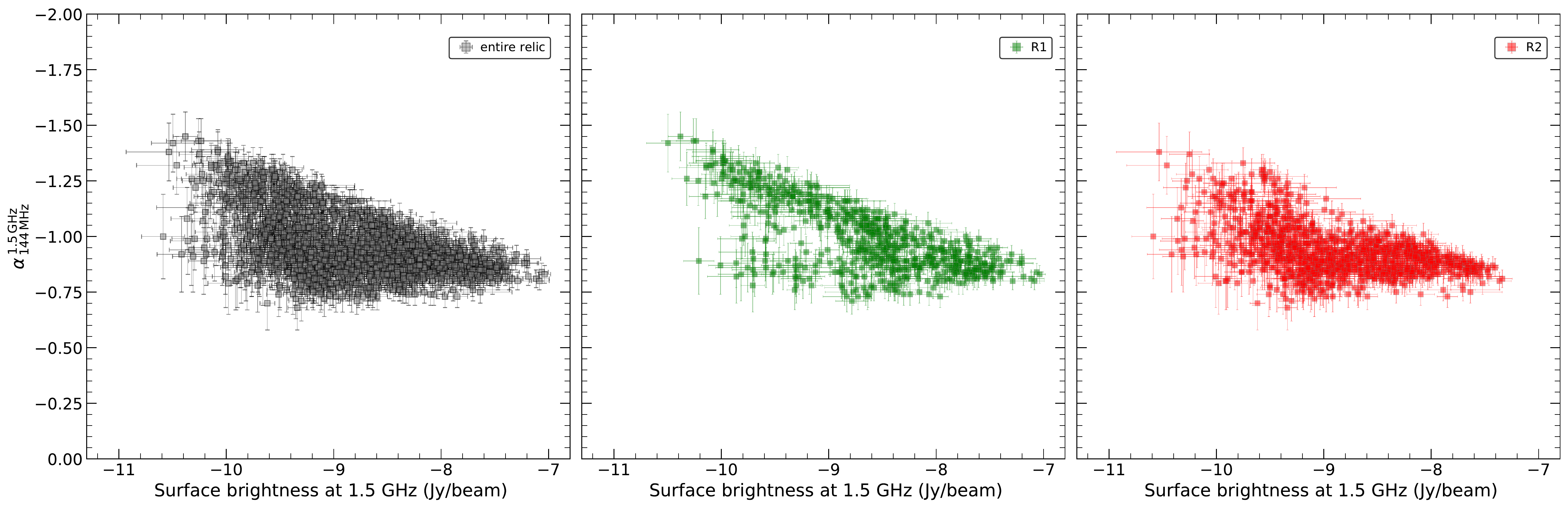}
\includegraphics[width=1.0\textwidth]{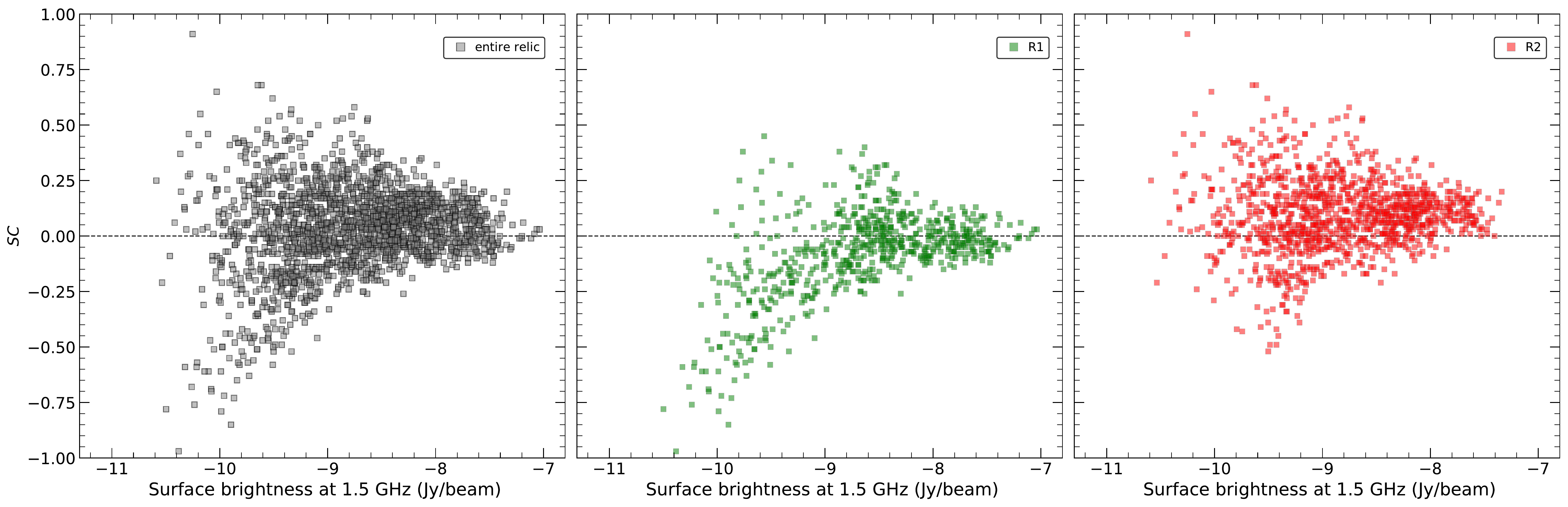}
 \caption{Radio surface brightness ($I_{\rm R}$) at 1.5\,GHz (in log scale) versus spectral index (top) and curvature (bottom) distributions across the relic. The surface brightness values were extracted from the same square boxes used in Fig.\ref{cc_plots1}  (i.e., $10\arcsec$ width, similar to the beam size). It is evident that structures with high radio surface brightness have flat spectral indices and show little curvature, implying that those regions are sites of acceleration (injection) and very likely trace high Mach number shocks.}
\label{brightness_vs_index}
\end{figure*}  

The morphology of radio relics can be understood by comparing the color-color diagrams with spectral aging models, such as the Jaffe-Perola \citep[JP;][]{Jaffe1973} and KGJP \citep[][]{Komissarov1994} models, see discussion in \cite{Rajpurohit2020a}. In addition, color-color diagrams are also sensitive to the relic's viewing angle \citep{Rajpurohit2021a}. If a relic is observed close to edge-on and a CRe population is injected with a uniform Mach number, the spectral shape is expected to follow the JP model. In such cases, the radio plasma ages in the downstream regions of the shock front, therefore a spectrum of a single spectral age is observed for each line of sight. 

If the shock front is instead inclined with respect to the line of sight, different ages are present along the line of sight and thus contribute to the observed spectrum. In this case, the spectral shape can be described by the KGJP model. Simulations also suggest that if a relic is seen face-on, a broad range of spectral index values is expected in the observed color-color diagram \citep{Rajpurohit2021a}. The observed spectral shape of the entire Abell 2256 relic is clearly inconsistent with the JP and KGJP aging models. However, a part of R1 can be explained by the KGJP model, see Fig.\,\ref{cc_plots1} (middle panel). The distribution of curvature seems to provide convincing evidence that the southern part of the relic in Abell 2256 is seen nearly face-on (i.e., R2) but there is also the northern part, which is rather close to edge-on (i.e., R1).

Overall, the complex spectral shape and a significant spread in the color-color diagram for the relic suggest inhomogeneity in the magnetic field, Mach numbers and may be overlapping of substructures along the same line of sight. The spectral indices and curvature trends found for the Abell 2256 relic can be best understood by assuming a part of the shock front is seen close to face-on and the other edge-on. The spectral properties are in line with both the acceleration due to a shock crossing a turbulent ICM and re-acceleration (with the shock still crossing the cloud of pre-existing plasma).

\subsection{Spectral tomography}
\label{tomographysection}
We construct images using the ``spectral tomography" technique, first introduced by \cite{KatzStone1997}. This technique allows us to investigate local spectral index variations of overlapping features. The first application of this technique to relics revealed overlapping features along the line of sight for the relic in MACS\,J0717.5+3745 \citep{Rajpurohit2021a}. In this technique, one of the images scaled at a spectral index of $\alpha_{t}$, is subtracted from the other image as
\begin{equation}
I(\alpha_t)=I_{\nu_{1}}- \left(\frac{\nu_1}{\nu_2}\right)^{\alpha_{t}} I_{\nu_2},
\end{equation}
where ${ \nu_{1}} = 144\,\rm MHz$ and ${ \nu_{2}}=1.5\,\rm GHz$. In the resulting images, features with spectral index $\alpha_{t}$ will vanish because the spectral index approaches the true value. Features with spectral indices different than $\alpha_{t}$ will appear as regions of positive flux (light) or negative flux (dark). The differences in the images allow us to identify distinct features according to their spectral indices. 

\begin{figure*}[!thbp]
\centering
\includegraphics[width=1.00\textwidth]{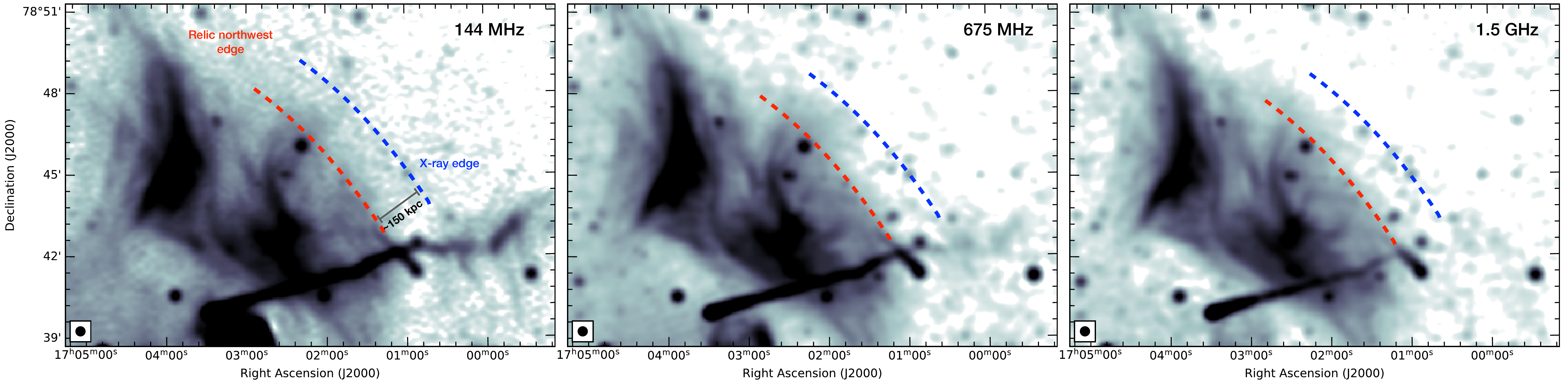}
 \caption{LOFAR, uGMRT Band\,4, and VLA 1.5\,GHz images at 20\arcsec resolution, revealing fainter radio emission to the northwest edge of the relic. A red dashed curve outlines the bright northwest edge of the relic emission detected in radio. The shock front identified by \cite{Ge2020} via X-ray observations is marked with a blue dashed curve.}
\label{offset}
\end{figure*}

The resulting images are shown in Fig.\,\ref{tomography}. The tomography images reveal that the relic is composed of multiple overlapping filaments. Although the individual filaments are quite distinct, it is clear that the spectral index differs between the brighter and fainter filaments. As shown in Fig.\,\ref{tomography} panel a, the prominent bright filaments appear dark, implying that they have spectral indices flatter than $-0.85$. Most of the filaments visible in the total power maps are recovered at $\alpha_{t}=-1.00$. This indicates that that the large-scale emission is rather dominated by filaments with spectral indices flatter than $-1.00$. This behavior is expected from high Mach number patches or regions that could be either part of the shock front, or other sites of fresh injection of CRe, causing the observed flatter spectral indices.

Moreover, the low surface brightness diffuse emission surrounding the filaments becomes clear at $\alpha_{t}=-1.25$ and $\alpha_{t}=-1.45$ (Fig.\,\ref{tomography} panels c and d). This suggests that the low surface brightness emission has a steeper spectral index than the filaments. Interestingly, the spectral index also varies across filaments, for example, the long filament shows the presence of some very dark patches at $\alpha_{t}=-1.25$ (Fig.\,\ref{tomography} panel c). In the region where the relic meets the long tail of the source C, we see clear evidence of several different filaments overlapping with the long tail, see Fig.\,\ref{tomography} panels b and c. 
 
Fig.\,\ref{tomography} panel c reveals that there are several filaments, particularly within the R2 region of the relic, crossing each other. The projection of different filamentary structures in turn may affect their spectral indices. Radio color-color diagrams obtained from simulations indicate that overlapping between different emitting structures within the relic results in concave spectra \citep{Rajpurohit2021a}. Therefore, the concave spectra seen in the radio color-color plots for the relic in Abell 2256 could be partially due to the projection of different filaments within the relic. 

\subsection{Radio brightness vs spectral index relation}
\label{SB_vs_index}

We examine how the spectral index and curvature are related to the radio surface brightness. The resulting scatter plots are shown in Fig.\,\ref{brightness_vs_index} top panels. The spectral index is calculated between 144 MHz and 1.5\,GHz while the radio surface brightness is taken from the high frequency. There is a positive correlation between the radio brightness and spectral index: brighter regions have flatter spectra.

The correlation between the brightness and spectral index is well defined in the northern part of the relic (Fig.\,\ref{brightness_vs_index} top middle panel). There are also some data points showing flat spectra but are fainter. Those points are located at the boundary of R1 and R2. 

Bright filaments are found to have flat spectral indices in the range $-0.7$ to $-0.9$ while low fainter diffuse emission shows a tendency toward steeper indices ($\alpha\geq-1.10$). Similar trends are reported by \cite{Owen2014}. The flat spectral indices in bright regions may hint that they trace the shock regions with the highest Mach numbers and thus the sites of fresh injection, dominating the production of relativistic electrons that are responsible for the observed radio emission.

In the bottom panels of Fig.\,\ref{brightness_vs_index}, we show the relation of radio surface brightness to spectral curvature. In general, the bright regions show little curvature. The regions of low surface brightness show significant curvature at R1.  On the other hand, R2 shows mostly concave spectral curvature for both high and low surface brightness regions. If bright regions trace the highest Mach number at the shock surface and the sites of fresh injection, indeed, we do not expect significant curvature. This seems to be consistent with the idea that bright filaments are shock-related structures. Numerical simulations also show a positive relation of the magnetic field and Mach number with the surface brightness \citep[e.g., see Fig.\,15 in ][]{Paola2021}. This again hints that at the sites of fresh injection, the higher the B-field and Mach number, the brighter the radio emission.

Interestingly, the boxes in the relic R2 generally show concavely shaped spectra, in clear contrast to R1. We speculate this difference reflects that R1 is seen rather edge-on, while R2 is seen more face-on. For the edge-on geometry, the surface brightness correlates with the distance to the site of CRe injection at the shock front. For this geometry, low surface brightness boxes are further downstream and show a steeper spectral index and show a more curved spectrum because of the aging of the electron energy distribution. For the face-on geometry, one would expect for all boxes a power-law spectrum identical to the overall spectrum. However, the majority of the spectra in the relic R2 are evidently concave. A possible reason for this spectral shape is that the magnetic field strength is higher than average in a region very close to the shock front. This enhanced field strengths close to the shock would cause concave spectra and a spectral index less steep than -1, as observed for most of the bright boxes in R2. When averaging over the entire relic R2, that is averaging over a large variation of Mach numbers and of magnetic field strengths, this `boost of the injection spectrum' is possibly averaged out and hence not noticeable in the overall spectrum. Thus, the concave spectra in R2 may point to an enhanced magnetic field strength in the shock region which decays quickly downstream of the shock.

\subsection{ Discrepancy and offset between radio and X-ray detected shocks}
\label{offset_dis}
From the radio integrated spectral index, we obtained a shock of $\mathcal{M}=5.4^{-1.0}_{-0.6}$. Recently, from X-ray observations, \cite{Ge2020} detected the presence of density and temperature discontinuities at the northwest edge of the relic in Abell 2256. The shock Mach number obtained from X-ray observations is much lower than that of the radio observations, namely $\mathcal{M}_{\rho}=1.23\pm0.06$ and $\mathcal{M}_{\rm T}=1.62\pm1.2$ (based on the density and temperature jumps, respectively). There is clearly a large discrepancy between the Mach numbers obtained from radio and X-ray observations. In several other radio relics, Mach numbers obtained from radio observations are significantly larger than the corresponding X-ray derived Mach numbers \citep[e.g.,][]{vanWeeren2016a}. 

Cosmological simulations suggest that the radio and X-ray structures projected to the sky may consist of shock surfaces of different Mach numbers: radio observations tend to pick the strongest Mach number shocks, while X-ray observations pick the low Mach number shocks along a given line of sight \citep{Skillman2013,Hong2015,Roh2019,Wittor2019,Paola2021}. Recently, \cite{Wittor2021} studied the radio versus X-ray Mach number discrepancy by comparing numerical simulations with the integrated radio spectra of the observed relics. They found that radio and X-ray Mach numbers differ intrinsically because they trace different parts of the underlying Mach number distribution. Their study suggests that the radio Mach number reflects the actual width of the underlying Mach number distribution in radio relics.

Our spectral index and curvature analysis implies that the southern part relic is seen close to face-on. As reported by \cite{Wittor2021}, the X-ray Mach number reflects the mean of the underlying Mach number distribution. However, the X-ray Mach number is very sensitive to the orientation of the relic and, hence, it mostly underestimates the mean. On the other hand, radio Mach numbers do not depend on the relic orientation and are always biased toward the strongest shocks \citep{Rajpurohit2021a,Wittor2021}. Therefore, the shock detected in radio observations may not be necessarily brightest in X-rays or vice versa. While in this relic, the large angle toward the observer makes it possible to have these structures clearly visible across the full shock extent, an X-ray observation can only detect a weaker shock through a standard jump condition analysis.

In Fig.\,\ref{offset}, we show a $20\arcsec$ resolution LOFAR 144\,MHz, uGMRT 675\,MHz and VLA 1.5\,GHz images. The LOFAR image reveals a much fainter radio emission to the northwest of the relic emission. This fainter emission is also seen in the uGMRT Band\,4 low resolution image. \cite{Ge2020} reported that there is an apparent offset of about 150\,kpc (projected) between the relic emission at 1.5\,GHz and the discontinuity detected via X-ray observations. However, in our VLA image, we do detect a part of this fainter emission (see Fig.\,\ref{offset} right panel) at $20\arcsec$. We do not find an offset as large as 150\,kpc in our 1.5\,GHz radio map. From the LOFAR 144\,MHz image, it is clear that the faint radio emission reaches the jump detected in X-ray observations very well. This emission is a factor of 5-10 fainter than the bright radio emission at the northwest edge of the relic. At all three frequencies, it is evident that there is faint radio emission between the location of the shock measured from X-ray observations and the radio relic edge used by \cite{Ge2020}. We argue that there is no offset between the radio relic and the X-ray-detected shock. 

\section{Summary and conclusions}
\label{summary}
In this work, we have presented the first deep wideband, high-resolution, low-frequency uGMRT and LOFAR radio observations of the famous galaxy cluster Abell 2256. Previous studies of the cluster at frequencies below 1\,GHz have been limited by their poor resolution and sparser \textit{uv}-coverage. Our new uGMRT observations and LOFAR (120-169\,MHz) in combination with the archival VLA L- and S-band (1-4\,GHz) observations, provide crucial insights into the origin of the large relic. We summarize our main findings as follows:

\begin{itemize} 
\item Our new images confirm the existence of complex filamentary structures in the Abell 2256 relic down to 300\,MHz. The relic emission appears more extended toward low frequencies. We also detect the large halo emission at the cluster center. 

\item Using our high sensitivity radio observations, we constrained the integrated radio spectrum of the radio relic. The overall spectra of the entire relic and subregions (R1 and R2) closely follow a single power law between 144\,MHz and 3\,GHz. We do not find any evidence of low frequency spectral steepening of the relic emission below 3\,GHz. 

\item For the relic, we find an integrated spectral index of $\alpha = -1.07\pm0.02$, which is significantly steeper than all previously reported values. Our findings indicate that the relic in Abell 2256 follows the stationary state shock condition like other well-known radio relics. Our analysis also suggesst that the fraction of energy channeled into the acceleration of suprathermal electrons is about 1\,\% or less for a magnetic field strength of a few $\mu$G in the relic region.

\item  The northern part of the relic (R1) shows clear gradients in both spectral index and spectral curvature. Small-scale fluctuations in the  spectral index are measured in the southern part of the relic (R2) without any hint of spectral gradient. The R2 shows a hint of concave curvature. The difference in the spectral index and curvature between R1 and R2 may be due to projection effects. 

\item The radio color-color diagrams reveal a complex curvature distribution with a broad range of concave, power-law, and convex spectra. This very likely implies inhomogeneity in the magnetic field and varying Mach number distributions across the relic structure. The observed spectral shape suggests that R1 is seen edge-on and R2 close to face-on.  

\item We find that there are several filamentary structures in the relic with different spectral indices crossing each other. The individual filaments show different spectral indices; bright filaments show a spectral index flatter than $-0.85$, while less bright filaments are $-0.85\leq\alpha\leq-1.00$.

\item A positive correlation is observed between the radio brightness and spectral index, so that the brighter regions/filaments have flatter spectra and show little curvature. This suggests that those filaments are the sites of fresh injection and trace high Mach number shock regions.  

\item In our 144\,MHz and 675\,MHz images, we detected low surface brightness emission to the west of the relic. This faint radio emission extends to the location of a shock detected in X-rays. Unlike previous claims, our finding suggests that there is no offset between the relic and the shock detected in X-rays.

\end{itemize} 

In conclusion, our new high-resolution and broadband campaign of observation yields a consistent view of the radio relic in Abell 2256. The complexities of the detected filament and substructures are understood in terms of a typical merger shock-wave propagation through a more turbulent, magnetized, and dynamically active ICM. We suggest that the relic surface underlines the distribution of Mach numbers. The highest values of this Mach distribution lead to flatter spectra of the injected electrons. As a consequence, the observed radio emission is dominated by the high tail of the Mach distribution, which dominates the injection of electrons and thus the observed radio emission. If the above picture is correct, the radio relic in Abell 2256 might not be particularly exceptional compared to most of the radio relics in the literature. However, through a favorable combination of inclination of the shock along the line of sight and timing of the merger event, only in Abell 2256  can we have such a clear view of the shock surface.

\section*{Acknowledgments}
KR, FV, and M. Brienza acknowledges financial support from the ERC Starting Grant ``MAGCOW" no. 714196. RJvW and A. Botteon acknowledge support from the VIDI research program with project number 639.042.729, which is financed by the Netherlands Organization for Scientific Research (NWO). D.W. is funded by the Deutsche Forschungsgemeinschaft (DFG, German Research Foundation) - 441694982. A. Bonafede, CJR, EB, and CS acknowledges support from the ERC through the grant ERC-Stg DRANOEL n. 714245.  W.F. acknowledges support from the Smithsonian Institution and the Chandra High Resolution Camera Project through NASA contract NAS8-03060. GB acknowledge PRIN INAF mainstream "Galaxy cluster science with LOFAR". MB and FdG acknowledges support from the Deutsche Forschungsgemeinschaft under Germany's Excellence Strategy - EXC 2121 ``Quantum Universe" -390833306. AD acknowledges support by the BMBF Verbundforschung under the grant 05A20STA. The GMRT is run by the National Centre for Radio Astrophysics (NCRA) of the Tata Institute of Fundamental Research (TIFR). LOFAR \citep{Haarlem2013} is the Low Frequency Array designed and constructed by ASTRON. It has observing, data processing, and data storage facilities in several countries, which are owned by various parties (each with their own funding sources), and that are collectively operated by the ILT foundation under a joint scientific policy. The ILT resources have benefited from the following recent major funding sources: CNRS-INSU, Observatoire de Paris and Universit\'{e} d'Orl\'{e}ans, France; BMBF, MIWF-NRW, MPG, Germany; Science Foundation Ireland (SFI), Department of Business, Enterprise and Innovation (DBEI), Ireland; NWO, The Netherlands; The Science and Technology Facilities Council, UK; Ministry of Science and Higher Education, Poland; The Istituto Nazionale di Astrofisica (INAF), Italy. This research made use of the LOFAR-UK computing facility located at the University of Hertfordshire and supported by STFC [ST/P000096/1], and of the LOFAR-IT computing infrastructure supported and operated by INAF, and by the Physics Dept. of Turin University (under the agreement with Consorzio Interuniversitario per la Fisica Spaziale) at the C3S Supercomputing Centre, Italy.

\bibliography{ref.bib}

\end{document}